%% file: main.tex
\pdfoutput=1
\documentclass[12pt]{article}
\usepackage[utf8]{inputenc}
\usepackage[height=8.85in,width=6.45in]{geometry}
\usepackage{hyperref}
\usepackage{todonotes}
\usepackage[most]{tcolorbox}

\usepackage{array}
\usepackage{multirow}
\usepackage{colortbl}
\usepackage[dvipsnames]{xcolor}
\usepackage{here}

\DeclareMathOperator{\Fer}{Fer}
\newcommand{\ABK}{\mathrm{ABK}}
\newcommand{\Pin}{\mathrm{Pin}}

\newtcolorbox[auto counter, number within=section]{statementbox}[1][]{
  colback=white, colframe=black,
  title=Statement~\thetcbcounter,
  fonttitle=\bfseries,
  boxrule=0.5pt,
  arc=2mm,
  left=6pt, right=6pt, top=6pt, bottom=6pt,
  #1
}

\newtcolorbox[auto counter, number within=section]{definitionbox}[1][]{
  colback=gray!10, colframe=gray!80,
  title=Definition~\thetcbcounter,
  fonttitle=\bfseries,
  boxrule=0.5pt,
  arc=2mm,
  left=6pt, right=6pt, top=6pt, bottom=6pt,
  #1
}

\newtcolorbox[auto counter, number within=section]{theorembox}[1][]{
  colback=gray!10, colframe=blue!80,
  title=Theorem~\thetcbcounter,
  fonttitle=\bfseries,
  boxrule=0.5pt,
  arc=2mm,
  left=6pt, right=6pt, top=6pt, bottom=6pt,
  #1
}

\newtcolorbox[auto counter, number within=section]{propositionbox}[1][]{
  colback=gray!10, colframe=green!60!black,
  title=Lemma~\thetcbcounter,
  fonttitle=\bfseries,
  boxrule=0.5pt,
  arc=2mm,
  left=6pt, right=6pt, top=6pt, bottom=6pt,
  #1
}

\newtcolorbox[auto counter, number within=section]{remarkbox}[1][]{
  colback=white, colframe=gray!50,
  title=Remark~\thetcbcounter,
  fonttitle=\bfseries,
  boxrule=0.3pt,
  arc=1.5mm,
  left=6pt, right=6pt, top=6pt, bottom=6pt,
  #1
}

\newtcolorbox{graybox}{breakable,
  colback=gray!40,     
  colframe=gray!15,    
  boxrule=0.2pt,       
  arc=0mm,             
  left=5pt, right=5pt, top=5pt, bottom=5pt
}

\newtcolorbox{lightgraybox}{breakable,
  colback=gray!7.5,     
  colframe=gray!15,    
  boxrule=0.2pt,       
  arc=0mm,             
  left=5pt, right=5pt, top=5pt, bottom=5pt
}

\input{head.tex}

\title{Web of dualities on non-orientable surfaces}

\author{Ippo Orii and Keita Tsuji}

\affil{}

\date{}
\begin{document}
    \begin{titlepage}
        \centering
        \begin{flushright}
        \end{flushright}
        \vskip 5em
        {\LARGE Web of dualities on non-orientable surfaces}
        \vskip 3em
        {\large Ippo Orii and Keita Tsuji}
        \vskip 1.5em
        {\large Kavli Institute for the Physics and Mathematics of the Universe, \\\vspace*{1mm}
University of Tokyo, Kashiwa, Chiba 277-8583, Japan}
        \vskip 3em
        \input{abstract}
    \end{titlepage}

    \clearpage
    \setcounter{tocdepth}{2}
    \tableofcontents

    \input{1.tex}
    \input{2.tex}
    \input{3.tex}
    \input{4.tex}
    \input{5.tex}
    \input{6.tex}

 \bibliographystyle{ytamsalpha}
 \def\arxivfont{\rm}
 \baselineskip=.95\baselineskip
\bibliography{ref}

\end{document}

%% file: head.tex
\usepackage{amsfonts}
\usepackage{amssymb}
\usepackage{amsmath}
\usepackage{amsthm}
\usepackage{mathtools}
\usepackage{cleveref}
\usepackage{url}
\usepackage{tikz}
\usepackage{tikz-cd}
\usepackage{adjustbox}
\usepackage{enumitem}
\usepackage{subcaption}
\usepackage{graphicx}
\usepackage{afterpage}
\usepackage{changepage}
\usepackage{authblk}
\usepackage{longtable}
\usepackage{bm}
\usepackage{multicol}
\usepackage{wrapfig}
\usepackage{enumitem}
\usepackage{braket}

\usetikzlibrary{knots}
\usetikzlibrary{calc}

\theoremstyle{plain}

\theoremstyle{definition}

\theoremstyle{remark}

\crefname{maintheorem}{Theorem}{Theorems}
\crefname{claim}{Claim}{Claims}

\newtheoremstyle{restated}
    {\topsep}{\topsep} 
    {\itshape}         
    {}                 
    {\bfseries}        
    {.}                
    { }                
    {\thmname{#1} \ref{#3} {\normalfont(Restated)}}
    
\theoremstyle{restated}
    \newtheorem{restate-theorem}{Theorem}
    \newtheorem{restate-proposition}{Proposition}
    \newtheorem{restate-corollary}{Corollary}
    
\numberwithin{equation}{section}
\numberwithin{table}{section}

\newcommand{\ZZ}{\mathbb{Z}}

\renewcommand{\epsilon}{\varepsilon}

\DeclareMathOperator{\Tr}{Tr}

\newcommand{\bZ}{\mathbb{Z}}


%% file: abstract.tex
\begin{abstract}
It is known that a two-dimensional bosonic theory with a non-anomalous $\mathbb{Z}_2$ symmetry can be fermionized. Recent work shows that if the bosonic theory also has non-anomalous time-reversal symmetry, fermionization extends to non-orientable surfaces and yields a fermionic theory that depends on a $\mathrm{Pin}^-$ structure. Besides fermionization, one can define various topological manipulations, such as gauging and stacking invertible phases, which together generate a web of dualities. We prove that their group structure is the dihedral group $D_8$ of order 16. Furthermore, we systematically investigate the web from two perspectives: Symmetry TFT and actions on sectors of the $S^1$ Hilbert space.
\end{abstract}

%% file: 1.tex
\section{Introduction and summary}
Symmetry plays an important role in quantum field theory and is one of the key
characteristics of a physical system. Given a theory with a symmetry, one can
consider various operations such as gauging the symmetry or stacking with an
invertible field theory. One of the interesting aspects of these operations is
that they sometimes exhibit non-trivial dualities between theories.

A well-known and beautiful example is a web of dualities in two dimensions,
which is also the main focus of this work. Suppose we are given a bosonic theory
with a non-anomalous \(\mathbb{Z}_2\) symmetry. There are two natural ways to obtain new theories from the original one.
One is to gauge the $\mathbb{Z}_2$ global symmetry, and the other is to fermionize
the theory.
Gauging the symmetry yields another bosonic theory equipped with a dual
$\mathbb{Z}_2$ symmetry.
In contrast, fermionization produces a fermionic theory with the fermion parity.
Remarkably, these two procedures are related by a duality.
Starting from the original bosonic theory, we first fermionize it and then stack
the resulting fermionic theory with the Arf invariant.
Upon subsequently bosonizing, we recover precisely the dual $\mathbb{Z}_2$
bosonic theory obtained by gauging the original theory.
This duality was studied in~\cite{Kapustin:2017jrc,Karch:2019lnn,Ji:2019ugf,Gaiotto:2020iye,Hsieh:2020uwb,Kulp:2020iet} and has since been extended in
several directions~\cite{Duan:2023ykn,Turzillo:2023yyr,Spieler:2025hyr}.

Such operations on a theory are naturally described from the perspective of
Symmetry TFT (SymTFT)~\cite{Apruzzi:2021nmk, Schafer-Nameki:2023jdn, Freed:2022qnc, Kaidi:2023maf}. SymTFT provides a framework for understanding these
operations in terms of topological manipulations in a one-dimension-higher
topological field theory. For example, consider a theory defined on a spacetime
\(\Sigma\). We may enlarge the spacetime to \(\Sigma \times [0,1]\) and assume
that the bulk theory is a topological field theory. From this point of view, the
original theory can be seen as a boundary condition associated with \(\Sigma\).
Moreover, a topological field theory may admit gapped (topological) boundary
conditions. In such cases, one can construct a gapped boundary condition which is 
topological in the sense that it can be freely deformed without changing the
topology. As a result, the interval can be shrunk, recovering the original
partition function as an inner product with these two boundary states (See Figure~\ref{Sym TFT}).

From this perspective, operations on a theory such as gauging are described as
the insertion of a topological defect along the interval (See Figure~\ref{Top manipulation in sym tft}). In this way,
symmetries and operations on the theory are encoded by a choice of gapped
boundary together with the insertion of a topological defect, resulting in a
topological manipulation of the theory.

Let us now recall the web of dualities mentioned above. Roughly speaking, the
key point of this duality is that one can fermionize a theory by coupling a
\(\mathbb{Z}_2\) gauge field to a \(\bZ_2\)-valued quadratic refinement, since quadratic
refinements are in one-to-one correspondence with spin structures on a surface.
In the non-orientable case, one can instead consider fermionization that depends
on a \(\mathrm{Pin}^-\) structure, using a \(\mathbb{Z}_4\)-valued quadratic
refinement, which is in one-to-one correspondence with \(\mathrm{Pin}^-\)
structures on a surface. It has been pointed out that the web of dualities
persists in such a setup~\cite{Turzillo:2023yyr,Spieler:2025hyr}. The minimal basics of quadratic refinements are summarized in the Appendix~\ref{app}.\footnote{In this paper, we restrict our attention to the \(\mathrm{Pin}^-\) case.
This is because
there is no quadratic-refinement–like object associated with
\(\mathrm{Pin}^+\) structures. Moreover, when dealing with general surfaces,
it is natural to focus on \(\mathrm{Pin}^-\) structures, since not all surfaces
admit \(\mathrm{Pin}^+\) structures.
}

In this paper, we elucidate that the group structure underlying the operations
appearing in the web of dualities is the dihedral group $D_8$ of order 16, and we interpret them from the Symmetry TFT perspective. Furthermore, we place the theory on a spatial circle \(S^1\) and analyze the resulting dualities in terms of sectors of the Hilbert space on the circle, illustrating them with an explicit example based on the Majorana/Ising CFT. The dualities among the sectors are summarized in Table~\ref{table:Summary}.

\begin{figure}[t]
  \centering
  \begin{tikzpicture}[
    scale=0.8,
    transform shape,
    font=\Large,
    >=Stealth,
    every node/.style={inner sep=2pt},
  ]

  \node (B)  at (0,4)  {$T_B$};
  \node (F)  at (8,4) {$T_F$};

  \node (Bt) at (0,0)  {$O^B(T_B)$};
  \node (Ft) at (8,0) {$O^F(T_F)$};

  \draw[<->, line width=0.9pt] (B) -- (F)
    node[midway, above=10pt] {\small $\Fer$: fermionization}
    node[midway, below=10pt] {\small $\Fer^{-1}$: bosonization};

  \draw[<->, line width=0.9pt] (B) -- (Bt)
    node[midway, anchor=east, xshift=-15pt] {\small $O^B$: gauging};

  \draw[<->, line width=0.9pt] (F) -- (Ft)
    node[midway, anchor=west, xshift=15pt] {\small $O^F$: stacking + shift};

  \draw[<->, line width=0.9pt] (Bt) -- (Ft)
    node[midway, above=10pt] {\small $\Fer$: fermionization}
    node[midway, below=10pt] {\small $\Fer^{-1}$: bosonization};

  \end{tikzpicture}

  \caption{A web of dualities between fermionization and gauging.}
  \label{Web of duality}
\end{figure}
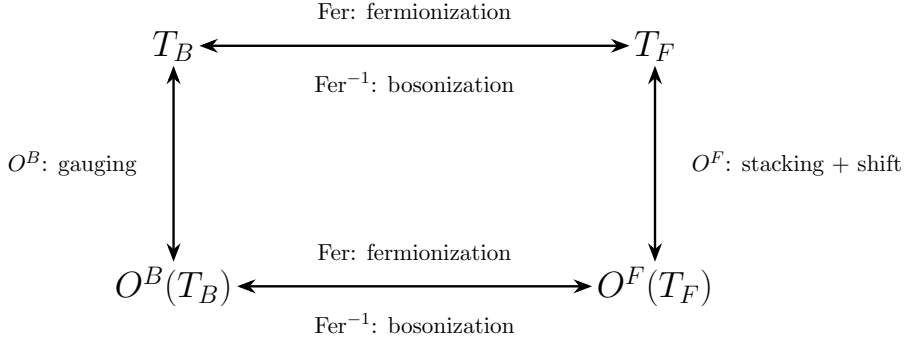

\paragraph{Organization of the paper}
In Section~2, we define topological manipulations such as fermionization and
gauging, and explain the web of dualities shown in Figure~\ref{Web of duality}.
In Section~3, we interpret the operations defined in Section~2 from the perspective
of Symmetry TFT.
In Section~4, we place a theory on a circle \(S^1\) and clarify the dualities
among different sectors of partition functions on the torus and the Klein
bottle.
In Section~5, we illustrate these dualities using the Majorana/Ising CFT as an
explicit example.
For convenience, we also review the basics of
\(\mathrm{Spin}/\mathrm{Pin}^-\) structures and
\(\mathbb{Z}_2/\mathbb{Z}_4\)-valued quadratic refinements, which are used
throughout the paper.

%% file: 2.tex
 \section{Topological manipulations} 
In this section, we define three topological manipulations $O^B,S_1^B,$ and $S_2^B$ on a time-reversal-symmetric bosonic theory with $\bZ_2$ symmetry and the corresponding manipulations $O^F,S_1^F,$ and $S_2^F$ on a fermionic theory, following \cite{Spieler:2025hyr}. We then investigate the group structure of these operations.

\subsection{Bosonic theory}

A time-reversal symmetric bosonic theory $T_B$ with a $\mathbb{Z}_2$ symmetry gives a partition function $Z_{T_B}[\Sigma;A]$ on a two-dimensional spacetime manifold $\Sigma$ and a background $\mathbb{Z}_2$ field $A\in H^1(\Sigma;\mathbb{Z}_2)$.\footnote{We assume that the $\ZZ_2$ symmetry and the time-reversal symmetry are non-anomalous, so $A$ is valued in $H^1(\Sigma;\ZZ_2)$ and $\Sigma$ may be non-orientable.} When $\Sigma$ is clear from the context, we abbreviate $Z_{T_B}[\Sigma;A]$ as $Z_{T_B}[A]$. We define three operations on $T_B$:
\begin{align}
    Z_{O^B(T_B)}[A] &= \frac{1}{|H^1(\Sigma;\ZZ_2)|^{1/2}} \sum_{a\in H^1(\Sigma;\mathbb{Z}_2)}(-1)^{\int a\cup A}Z_{T_B}[a],\label{op on bos theory O}\\
    Z_{S_1^B(T_B)}[A] &= Z_{T_B}[A](-1)^{\int A\cup w_1},\label{op on bos theory S_1}\\
    Z_{S_2^B(T_B)}[A] &= Z_{T_B}[A](-1)^{\int w_2},\label{op on bos theory S_2}
\end{align}
where $w_1,w_2$ are the first and second Stiefel--Whitney classes of $\Sigma$, respectively. $O^B$ is the usual $\mathbb{Z}_2$ gauging, $S_1^B$ is stacking the SPT $(-1)^{\int A\cup w_1}$, and $S_2^B$ is stacking the Haldane phase $(-1)^{\int w_2}$.

\subsection{Fermionic theory}

A time-reversal symmetric fermionic theory $T_F$ with $T^2=1$ gives a partition function $Z_{T_F}[\Sigma;\eta]$ on a spacetime manifold $\Sigma$ and a $\Pin^-$ structure $\eta$ on $\Sigma$. We can fermionize a bosonic theory $T_B$ in the following way:\footnote{Note that this operation becomes more complicated in dimensions higher than two. See~\cite{Gu:2012ib,Bhardwaj:2016clt} for orientable cases, and~\cite{Kobayashi:2019xxg,Kobayashi:2022qhc} for non-orientable cases.
To obtain a Pin$^{-}$ fermionic theory via fermionization, one must start from
a bosonic theory equipped with a $(d-2)$-form $\mathbb{Z}_2$ symmetry whose anomaly is
given by $(-1)^{\int \mathrm{Sq}^2(a_{d-1})}$, which is trivial in $d=2$.
Here $\mathrm{Sq}^2$ denotes the Steenrod square, and $a_{d-1}$ is the corresponding
$\mathbb{Z}_2$ background $(d-1)$-form gauge field.}
\begin{equation}
    Z_{\Fer(T_B)}[\eta] = \frac{1}{|H^1(\Sigma;\ZZ_2)|^{1/2}} \sum_{a\in H^1(\Sigma;\mathbb{Z}_2)} \exp\left(\frac{\pi i}{2}q_\eta (a)\right)Z_{T_B}[a],
\end{equation}
where $q_\eta$ is the quadratic refinement corresponding to a $\Pin^-$ structure $\eta$. Properties of quadratic refinements are summarized in the Appendix~\ref{Nonorientable abk}.

The inverse operation of fermionization is bosonization:
\begin{equation}
    Z_{\Fer^{-1}(T_F)}[A] = \frac{1}{|H^1(\Sigma;\ZZ_2)|^{1/2}} \sum_{\eta\in\Pin^-(\Sigma)}\exp\left(-\frac{\pi i}{2}q_\eta (A)\right)Z_{T_F}[\eta],
\end{equation}
where $\Pin^-(\Sigma)$ is the set of all $\Pin^-$ structures on $\Sigma$. $\Pin^-(\Sigma)$ is a torsor over $H^1(\Sigma;\ZZ_2)$. Here, we assume that the gravitational anomaly vanishes mod $16$ so that bosonization is equivalent to gauging fermion parity and well-defined.\footnote{
Since the set of spin structures forms a torsor over $H^{1}(\Sigma,\mathbb{Z}_2)$,
one might expect that gauging fermion parity produces a bosonic theory equivalent to summing over all spin structures.
However, there is a subtle but important distinction between these two procedures.
In fact, they are equivalent only when the gravitational anomaly vanishes mod $16$. See~\cite{BoyleSmith:2024qgx} for more discussion.}

Fermionization of the trivial theory $Z_\text{Tri}[A]=1$ is
\begin{equation}
    Z_{\ABK}[\eta]=\frac{1}{|H^1(\Sigma;\ZZ_2)|^{1/2}} \sum_a\exp\left(\frac{\pi i}{2}q_\eta (a)\right)=\exp\left(\frac{\pi i}{4}\ABK(\eta)\right),
\end{equation}
where $\ABK(\eta)$ is the $\mathbb{Z}_8$-valued invariant called the Arf-Brown-Kervaire (ABK) invariant.

We can define the operations corresponding to $O^B,S_1^B,$ and $S_2^B$ on a fermionic theory by similarity transformation:
\begin{equation}
    O^F = \Fer \circ\ O^B \circ \Fer^{-1},\quad
    S_1^F = \Fer \circ\ S_1^B \circ \Fer^{-1},\quad
    S_2^F = \Fer \circ\ S_2^B \circ \Fer^{-1}.
\end{equation}
This definition makes the diagrams of dualities commute; Figure~\ref{Web of duality} is an example for $O^B,O^F$.

To calculate the explicit expressions of these operations, the following relation is useful:\footnote{This relation follows from the property of the Steenrod square $\mathrm{Sq}^1(A) = A \cup A = A \cup w_1$. See e.g.~\cite[Appendix~3]{Kapustin:2014gma}.\label{a^2=aw_1}}
\begin{equation}
    A\cup A=A\cup w_1\quad\text{for all}\ A\in H^1(\Sigma;\ZZ_2).
\end{equation}

The expression for $S_1^F$ is\footnote{Our convention of fermionization is defined so that the expression for $S_1^F$ is simple.}
\begin{equation}
    \begin{aligned}
        Z_{S_1^F(T_F)}[\eta] &= Z_{\Fer(S_1^B(\Fer^{-1}(T_F)))}[\eta]\\
        &=\frac{1}{|H^1(\Sigma;\ZZ_2)|}\sum_{a,\eta'}\exp\left(\frac{\pi i}{2}q_\eta(a)\right)(-1)^{\int a\cup w_1}\exp\left(-\frac{\pi i}{2}q_{\eta'}(a)\right)Z_{T_F}[\eta']\\
        &=\frac{1}{|H^1(\Sigma;\ZZ_2)|}\sum_{a,b}\exp\left(\frac{\pi i}{2}q_\eta(a)\right)(-1)^{\int a\cup w_1}\exp\left(-\frac{\pi i}{2}q_{\eta+b}(a)\right)Z_{T_F}[\eta+b]\\
        &=\frac{1}{|H^1(\Sigma;\ZZ_2)|}\sum_{a,b}(-1)^{\int a\cup (b+w_1)}Z_{T_F}[\eta+b]\\
        &=Z_{T_F}[\eta+w_1].
    \end{aligned}
\end{equation}
The third equality comes from changing a sum over all $\Pin^-$ structures into a sum over $H^1(\Sigma;\ZZ_2)$. The fourth equality uses $q_{\eta+b}(a)=q_\eta(a)+2\int a\cup b$.

Next, the expression for $O^F$ is
\begin{equation}
    \begin{aligned}
        Z_{O^F(T_F)}[\eta] &= Z_{\Fer(O^B(\Fer^{-1}(T_F)))}[\eta]\\
        &=\frac{1}{|H^1(\Sigma;\ZZ_2)|^{3/2}}\sum_{a,b,\eta'}\exp\left(\frac{\pi i}{2}q_\eta(a)\right)(-1)^{\int a\cup b}\exp\left(-\frac{\pi i}{2}q_{\eta'}(b)\right)Z_{T_F}[\eta']\\
        &=\frac{1}{|H^1(\Sigma;\ZZ_2)|^{3/2}}\sum_{a,b,c}\exp\left(\frac{\pi i}{2}q_\eta(a)\right)(-1)^{\int a\cup b}\exp\left(-\frac{\pi i}{2}q_{\eta+c}(b)\right)Z_{T_F}[\eta+c]\\
        &=\frac{1}{|H^1(\Sigma;\ZZ_2)|^{3/2}}\sum_{a,b,c}\exp\left(\frac{\pi i}{2}q_\eta(a+b)\right)(-1)^{\int (w_1+c)\cup b}Z_{T_F}[\eta+c]\\
        &=Z_{\ABK}[\eta]\frac{1}{|H^1(\Sigma;\ZZ_2)|}\sum_{b,c}(-1)^{\int (w_1+c)\cup b}Z_{T_F}[\eta+c]\\
        &=Z_{\ABK}[\eta]Z_{T_F}[\eta+w_1],
    \end{aligned}
\end{equation}
where we used
\begin{equation}
    \begin{aligned}
        q_\eta(a)-q_{\eta+c}(b)&=q_\eta(a)-q_\eta(b)-2\int b\cup c\\
        &=q_\eta(a+b)+2\int (a+b)\cup b+2\int b\cup c\\
        &=q_\eta(a+b)+2\int (a+w_1+c)\cup b.
    \end{aligned}
\end{equation}

Finally, the expression for $S_2^F$ is the same as in a bosonic theory. In summary, the topological manipulations on a fermionic theory are
\begin{align}
    Z_{O^F(T_F)}[\eta] &= Z_{T_F}[\eta+w_1]Z_{\ABK}[\eta],\\
    Z_{S_1^F(T_F)}[\eta] &= Z_{T_F}[\eta+w_1],\\
    Z_{S_2^F(T_F)}[\eta] &= Z_{T_F}[\eta](-1)^{\int w_2}.
\end{align}

\subsection{Group structure of manipulations}
The manipulations $O^B,S_1^B,$ and $S_2^B$ satisfy the relations:
\begin{equation}
    (O^B)^2=(S_1^B)^2=(S_2^B)^2=1,\quad S_2^B=(O^BS_1^B)^4,
\end{equation}
where the group structure is defined as $AB=A\circ B$. The first relation is obvious. Interestingly, the second relation becomes clear by fermionization. $O^F S_1^F$ is stacking ABK, and the fact $4\ABK(\eta)=\int w_2$ implies that stacking ABK four times yields $S_2^F$.

Thus, the independent relations are
\begin{equation}
    (O^B)^2=(S_1^B)^2=1,\quad (O^BS_1^B)^8=1.
\end{equation}
If there are no additional relations, this means that the operations $O^B,S_1^B,$ and $S_2^B$ form the dihedral group $D_8$ of order 16:
\begin{equation}
    D_8=\braket{s,t\mid s^2=t^2=1,(st)^8=1},
\end{equation}
whose 16 distinct elements are
\begin{equation}
    (O^B S_1^B)^n,S_1^B (O^BS_1^B)^n,\quad n=0,1,\dots,7.
\end{equation}

To see that these 16 elements are actually distinct, it is sufficient to check that they act on a theory in different ways. This can be easily seen from fermionization. For the trivial fermionic theory $Z_{\text{Tri}_F}[\eta]=1$,
\begin{align}
    Z_{(O^FS_1^F)^n(\text{Tri}_F)}[\eta]&=\exp\left(\frac{n\pi i}{4}\ABK(\eta)\right),\\
    Z_{S^F_1(O^FS_1^F)^n(\text{Tri}_F)}[\eta]&=\exp\left(\frac{n\pi i}{4}\ABK(\eta+w_1)\right)=\exp\left(-\frac{n\pi i}{4}\ABK(\eta)\right),
\end{align}
where we used $\ABK(\eta+w_1)=-\ABK(\eta)$ in $\ZZ_8$. This relation is easily seen from $Z_\ABK[\eta]=Z_{O^F S_1^F(\text{Tri}_F)}[\eta]=Z_{O^F(\text{Tri}_F)}[\eta]$ and
\begin{equation}
    1=Z_{\text{Tri}_F}[\eta]=Z_{O^F(\ABK)}[\eta]=\exp\left(\frac{\pi i}{4}\ABK(\eta)\right)\exp\left(\frac{\pi i}{4}\ABK(\eta+w_1)\right).
\end{equation}
Thus, $(O^FS_1^F)^n,S_1(O^F S_1^F),n>0$ change $\text{Tri}_F$ into non-trivial theories. Since $S_1^F$ is clearly different from $1$, the 16 elements are all distinct. We conclude that the group of $O^B,S_1^B,$ and $S_2^B$ is $D_8$.

\begin{figure}[H]
    \centering
    \begin{tikzpicture}[auto, scale=1, baseline=(current bounding box.center)]
        \foreach \t in {0,...,7} {
            \coordinate (p\t) at ({1.8 * sin(\t * 45 - 22.5)},{1.8 * cos(\t * 45 - 22.5)});
            \coordinate (P\t) at ({3.2 * sin(-\t * 45)},{3.2 * cos(-\t * 45)});
        }
        \foreach \t in {0,...,7} {
            \fill[black!25] (p\t) circle (0.25);
            \node at (p\t) {$\t$};
            \fill[black!25] (P\t) circle (0.25);
            \node at (P\t) {$\t'$};
            \pgfmathtruncatemacro{\tm}{mod(8-\t, 8)}
            \draw[stealth-stealth,blue,thick,dashed] ($(p\t)!0.2!(P\tm)$) -- ($(p\t)!0.8!(P\tm)$);
            \pgfmathtruncatemacro{\tp}{mod(\t+1, 8)}
            \draw[-stealth,thick] ($(p\t)!0.2!(p\tp)$) -- ($(p\t)!0.8!(p\tp)$);
            \draw[-stealth,thick] ($(P\t)!0.2!(P\tp)$) -- ($(P\t)!0.8!(P\tp)$);
            \pgfmathtruncatemacro{\tg}{mod(9-\t, 8)}
            \draw[stealth-stealth,red,thick] ($(p\t)!0.2!(P\tg)$) -- ($(p\t)!0.8!(P\tg)$);
        }
        \draw[stealth-stealth,thick,dotted] ($(p2)!0.1!(p6)$) -- ($(p2)!0.9!(p6)$);
        \node at (0.3,-0.3) {$S_2^B$};
        \node[red] at (0.7,2.5) {$O^B$};
        \node[blue] at (-0.7,2.5) {$S_1^B$};
        \node at (0,1.2) {$O^BS_1^B$};
    \end{tikzpicture}
    \hspace{2mm}
    \begin{tikzpicture}[auto, scale=1, baseline=(current bounding box.center)]
        \draw[-stealth,ultra thick] (-0.6, 0.5) -- (0.6, 0.5);
        \node at (0, 0.8) {$\Fer$};
        \draw[-stealth,ultra thick] (0.6, -0.5) -- (-0.6, -0.5);
        \node at (0, -0.8) {$\Fer^{-1}$};
    \end{tikzpicture}
    \hspace{2mm}
    \begin{tikzpicture}[auto, scale=1, baseline=(current bounding box.center)]
        \foreach \t in {0,...,7} {
            \coordinate (p\t) at ({1.8 * sin(\t * 45 - 22.5)},{1.8 * cos(\t * 45 - 22.5)});
            \coordinate (P\t) at ({3.2 * sin(-\t * 45)},{3.2 * cos(-\t * 45)});
        }
        \foreach \t in {0,...,7} {
            \fill[black!25] (p\t) circle (0.25);
            \node at (p\t) {$\t$};
            \fill[black!25] (P\t) circle (0.25);
            \node at (P\t) {$\t'$};
            \pgfmathtruncatemacro{\tm}{mod(8-\t, 8)}
            \draw[stealth-stealth,blue,thick,dashed] ($(p\t)!0.2!(P\tm)$) -- ($(p\t)!0.8!(P\tm)$);
            \pgfmathtruncatemacro{\tp}{mod(\t+1, 8)}
            \draw[-stealth,thick] ($(p\t)!0.2!(p\tp)$) -- ($(p\t)!0.8!(p\tp)$);
            \draw[-stealth,thick] ($(P\t)!0.2!(P\tp)$) -- ($(P\t)!0.8!(P\tp)$);
            \pgfmathtruncatemacro{\tg}{mod(9-\t, 8)}
            \draw[stealth-stealth,red,thick] ($(p\t)!0.2!(P\tg)$) -- ($(p\t)!0.8!(P\tg)$);
        }
        \draw[stealth-stealth,thick,dotted] ($(p2)!0.1!(p6)$) -- ($(p2)!0.9!(p6)$);
        \node at (0.3,-0.3) {$S_2^F$};
        \node[red] at (0.7,2.5) {$O^F$};
        \node[blue] at (-0.7,2.5) {$S_1^F$};
        \node at (0,1.2) {$O^FS_1^F$};
    \end{tikzpicture}
    \caption{The generic orbit of $D_8$. In a fermionic theory, $n,n'$ denote $T_{F,n},T'_{F,n}$, respectively.}
    \label{fig:generic_orbit}
\end{figure}
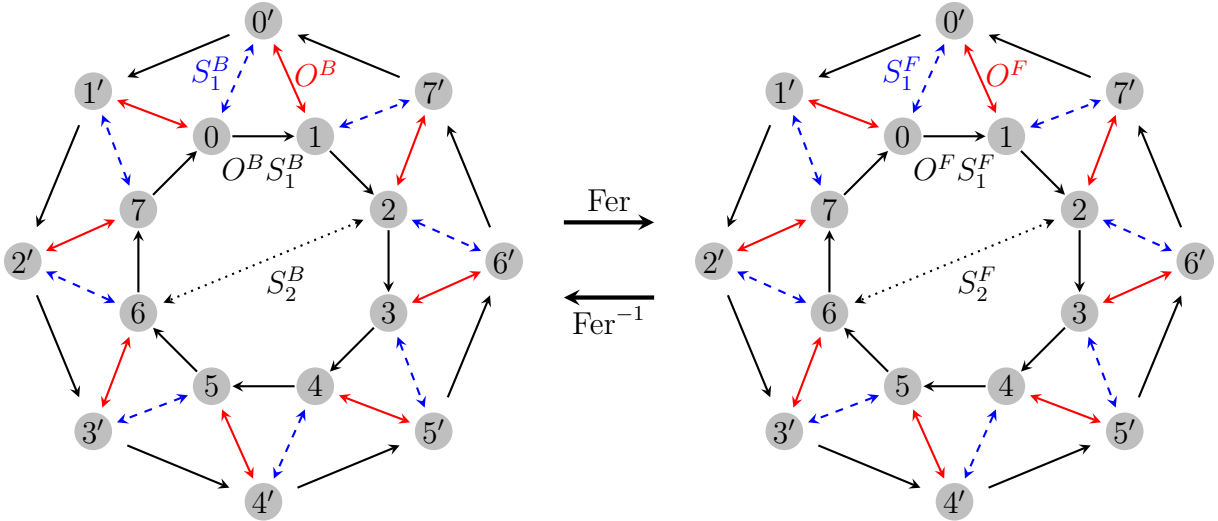

In general, each of the $16$ elements of $D_8$ maps a fermionic theory $T_F$ to a distinct theory. We write $(O^F S_1^F)^n(T_F)$ as $T_{F,n}$ and $S_1^F(O^FS_1^F)^n(T_F)$ as $T'_{F,n}$. Then the group action takes the form:
\begin{align}
    O^F(T_{F,n})&=T'_{F,-n+1},&O^F(T'_{F,n})&=T_{F,-n+1}\\
    S_1^F(T_{F,n})&=T'_{F,-n},&S_1^F(T'_{F,n})&=T_{F,-n},\\
    O^F S_1^F (T_{F,n})&=T_{F,n+1},&O^F S_1^F (T'_{F,n})&=T'_{F,n+1},\\
    S_2^F (T_{F,n})&=T_{F,n+4},&S_2^F (T'_{F,n})&=T'_{F,n+4}.
\end{align}
Figure~\ref{fig:generic_orbit} illustrates the generic orbit of $D_8$ as a web.

%% file: 3.tex
\section{Symmetry TFT}
\subsection{Bosonic theory}
Let us recall that we have three operations $O^B,S_1^B,$ and $S_2^B$ acting on
bosonic theories, as defined in~\eqref{op on bos theory O},\eqref{op on bos theory S_1}, and~\eqref{op on bos theory S_2}.
We now interpret these operations from the SymTFT perspective.
Given a two-dimensional bosonic theory with a non-anomalous $\mathbb{Z}_2$ symmetry, the associated SymTFT is the toric code $\mathbb{Z}_2 \times \mathbb{Z}_2$, namely
\(
\mathcal{Z}\bigl(\mathrm{Vec}_{\mathbb{Z}_2}\bigr).
\)
Such a theory can be realized as a BF theory with action
\begin{equation}
  S_{\mathrm{BF}} := \frac{2\pi}{2} \int_{\Sigma\times [0,1]} A_1 \cup \delta A_2 ,
\end{equation}
where $A_1$ and $A_2$ are $\bZ_2$ gauge fields.
This theory admits four types of topological line operators, three of which are nontrivial and are given by
\begin{equation}
  E(\Gamma) := \exp\!\left( \pi i \int_\Gamma A_1 \right), \quad
  M(\Gamma) := \exp\!\left( \pi i \int_\Gamma A_2 \right), \quad
  F(\Gamma) := E(\Gamma)\, M(\Gamma),
\end{equation}
where $\Gamma \in H_1(\Sigma \times [0,1]; \bZ_2)$.
Using Poincaré duality, we write
\(
E(\gamma) := E(\mathrm{PD}(\gamma)) = E(\Gamma),
\)
and hence regard $\gamma \in H^1(\Sigma; \bZ_2)$, since we will be interested in the action on the boundary.
In other words, we write $E(a)$ with $a \in H^1(\Sigma; \bZ_2)$, and similarly for $M(a)$ and $F(a)$. The fusion rules of bosonic line operators are
\begin{equation}
  E(a)^2 = M(a)^2=1.
\end{equation}
We obviously have the following relations:
\begin{equation}
       E(a)E(b)=E(a+b),\qquad
    M(a)M(b)=M(a+b).\label{addition of e m}
\end{equation}
In addition, there is a non-trivial relation:~\cite{Gaiotto:2014kfa}
\begin{equation}
  E(a) M(b)
  = (-1)^{\int a \cup b}\, M(b) E(a),
  \quad a,b \in H^1(\Sigma;\mathbb{Z}_2).\label{commutation E and B}
\end{equation}
Now recall that the choice of gapped boundary conditions corresponds to the
choice of a Lagrangian subgroup.\footnote{Here, we consider a TQFT in which the parity operation
acts trivially for the bulk theory.
In general, when dealing with non-orientable theories, the theory naturally
encodes a time-reversal or parity action on the anyons \(a\in \mathcal{C}\), arising
from orientation reversal of spacetime,
\(
a \;\mapsto\; \mathsf{T}a .
\)
In such situations, it is not always possible to span \(V(\Sigma)\) using
elements labeled by \(H^1(\Sigma,\mathbb{Z}_2)\), as we explain below.

As an example, consider the toric code
\(\mathcal{C}=\{1,e,m,f\}\), and suppose that \(\mathsf{T}\) exchanges
\(e \leftrightarrow m\).
Then the Hilbert space on the Klein bottle \(K\) is known to have dimension
\(
\dim V(K)
= \bigl|\{a \in \mathcal{C} \mid a = \overline{\mathsf{T}a}\}\bigr|
= 2
\neq
|H^1(K;\mathbb{Z}_2)| = 4 ,
\) with a subtlety in the notation.
In this sense, our setup corresponds to choosing a toric code in which
\(\mathsf{T}\) acts trivially,
\(
\mathsf{T}e = e,\,
\mathsf{T}m = m .
\)
See, for example~\cite{Barkeshli:2016mew,Orii:2025ktu}.
}
 In the present case, there are two even
Lagrangian subgroups and one odd Lagrangian subgroup, generated by \(E(a)\), \(M(a)\), and \(F(a)\),
respectively. Note that the odd Lagrangian subgroup generated by \(F(a)\)
corresponds to fermionic boundary conditions, which we discuss in the next subsection.

\paragraph{Boundary condition}
Let us condense the \(\mathbb{Z}_2\) subgroup generated by \(E(a)\) on
\(\Sigma \times \{0\} \cong \Sigma\).
This means that we consider a boundary condition \(\ket{e,0} \in V(\Sigma)\) satisfying
\begin{equation}
  E(a)\ket{e,0} = \ket{e,0}
  \quad
  \text{for all } a \in H^1(\Sigma;\mathbb{Z}_2).
\end{equation}
Then we easily see that
\(\ket{e,b} := M(b)\ket{e,0}\) is also an eigenstate of \(E(a)\), since
\begin{equation}
\begin{aligned}
  E(a)\ket{e,b}
  &= E(a) M(b)\ket{e,0} \\
  &= (-1)^{\int a \cup b} M(b) E(a)\ket{e,0} \\
  &= (-1)^{\int a \cup b} M(b)\ket{e,0} \\
  &= (-1)^{\int a \cup b}\ket{e,b}.
\end{aligned}
\end{equation}
Then one can check  that
\begin{equation}
    M(a)\ket{e,b}=\ket{e,a+b}.
\end{equation}
These states span the vector space associated with \(\Sigma\), namely,
\begin{equation}
  V(\Sigma)
  = \mathrm{span}_{\mathbb{C}}
  \bigl\{ \ket{e,a} \mid a \in H^1(\Sigma;\mathbb{Z}_2) \bigr\}.
\end{equation}
Similarly, we can construct another basis,
\begin{equation}
  V(\Sigma)
  = \mathrm{span}_{\mathbb{C}}
  \bigl\{ \ket{m,a} \mid a \in H^1(\Sigma;\mathbb{Z}_2) \bigr\},
\end{equation}
where the reference state \(\ket{m,0}\) satisfies
\begin{equation}
  M(a)\ket{m,0} = \ket{m,0}
  \qquad
  \text{for all } a \in H^1(\Sigma;\mathbb{Z}_2),
\end{equation}
and
\begin{equation}
  \ket{m,a} := E(a)\ket{m,0}.
\end{equation}
These define two gapped boundaries constructed from bosonic
\(\mathbb{Z}_2\) subgroups.
We fix the normalization by
\begin{align}
    \langle m,0|e,0 \rangle &= 1,\\
    \langle e,a|e,b \rangle &= |H^1(\Sigma;\mathbb{Z}_2)|^{1/2}\delta_{ab},\\
     \langle m,a|m,b \rangle &= |H^1(\Sigma;\mathbb{Z}_2)|^{1/2}\delta_{ab},
\end{align}
\begin{align}
    1&=\frac{1}{|H^1(\Sigma;\mathbb{Z}_2)|^{1/2}}\sum_{a\in H^1(\Sigma;\mathbb{Z}_2)} \ket{e,a}\bra{e,a},\\
    1&=\frac{1}{|H^1(\Sigma;\mathbb{Z}_2)|^{1/2}}\sum_{a\in H^1(\Sigma;\mathbb{Z}_2)} \ket{m,a}\bra{m,a},\label{identity m}
\end{align}
Then we have
\begin{equation}
\begin{aligned}
  \langle m,a | e,b \rangle
  &= \langle m,0 | E(a) M(b)| e,0 \rangle \\
  &= (-1)^{\int a \cup b}
     \langle m,0 |M(b) E(a)| e,0 \rangle \\
  &= (-1)^{\int a \cup b}
     \langle m,0 | e,0 \rangle \\
  &= (-1)^{\int a \cup b}.
\end{aligned}
\end{equation}
This implies that we can write
\begin{equation}
    \ket{e,a}=\frac{1}{|H^1(\Sigma;\bZ_2)|^{1/2}}\sum_{b\in H^1(\Sigma;\bZ_2)}(-1)^{\int a\cup b}\ket{m,b}.
\end{equation}

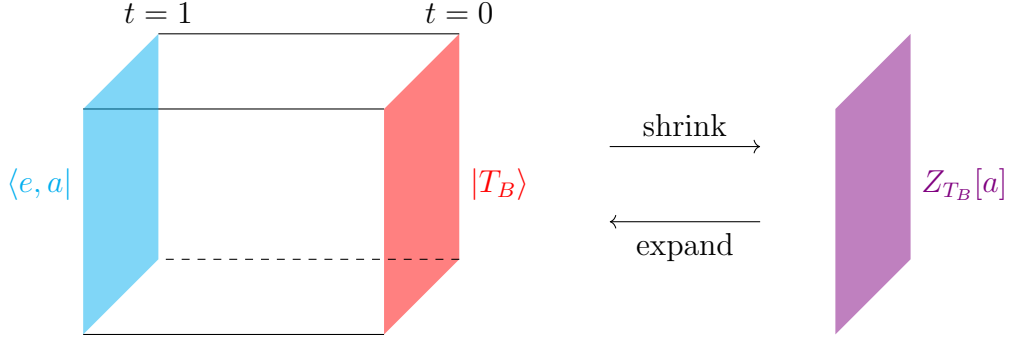
\begin{figure}[H]
    \centering
    \begin{tikzpicture}[auto, scale=1]
        \fill[red,opacity=0.5] (0,0) -- (1,1) -- (1,4) -- (0,3) -- cycle;
        \draw (0,0) -- (-4,0);
        \draw (0,3) -- (-4,3);
        \draw[dashed] (1,1) -- (-3,1);
        \draw (1,4) node[above] {$t=0$} -- (-3,4) node[above] {$t=1$};
        \fill[cyan,opacity=0.5] (-4,0) -- (-3,1) -- (-3,4) -- (-4,3) -- cycle;
        \node[right] at (1,2) {$\textcolor{red}{\ket{T_B}}$};
        \node[left] at (-4,2) {$\textcolor{cyan}{\bra{e,a}}$};

        \draw[->] (3,2.5) --node[midway, above] {shrink} (5,2.5);
        \draw[->] (5,1.5) --node[midway, below] {expand} (3,1.5);

        \fill[violet,opacity=0.5] (6,0) -- (7,1) -- (7,4) -- (6,3) -- cycle;
        \node[right] at (7,2) {$\textcolor{violet}{Z_{T_B}[a]}$};
    \end{tikzpicture}
    \caption{Symmetry TFT provides a framework for understanding the partition functions \(Z_{T_B}[a]\) as inner products between a physical (dynamical)
boundary state \(\ket{T_B}\) and a gapped (topological) boundary state
\(\bra{e,a}\), possibly with the insertion of a domain wall defect along the interval.}
    \label{Sym TFT}
\end{figure}

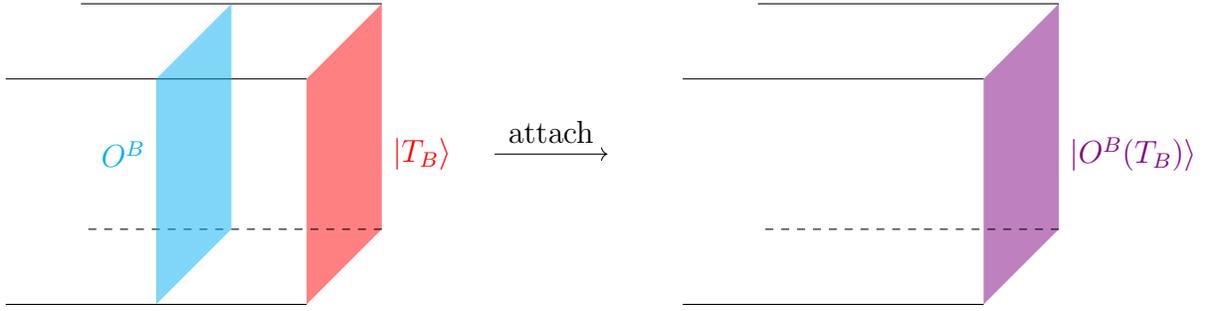
\begin{figure}[H]
    \centering
    \begin{tikzpicture}[auto, scale=1]
        \fill[red,opacity=0.5] (0,0) -- (1,1) -- (1,4) -- (0,3) -- cycle;
        \draw (0,0) -- (-4,0);
        \draw (0,3) -- (-4,3);
        \draw[dashed] (1,1) -- (-3,1);
        \draw (1,4) -- (-3,4);
        \fill[cyan,opacity=0.5] (-2,0) -- (-1,1) -- (-1,4) -- (-2,3) -- cycle;
        \node[right] at (1,2) {$\textcolor{red}{\ket{T_B}}$};
        \node[left] at (-2,2) {$\textcolor{cyan}{O^B}$};

        \draw[->] (2.5,2) --node[midway, above] {attach} (4,2);

        \fill[violet,opacity=0.5] (9,0) -- (10,1) -- (10,4) -- (9,3) -- cycle;
        \draw (9,0) -- (5,0);
        \draw (9,3) -- (5,3);
        \draw[dashed] (10,1) -- (6,1);
        \draw (10,4) -- (6,4);
        \node[right] at (10,2) {$\textcolor{violet}{\ket{O^B(T_B)}}$};
    \end{tikzpicture}
    \caption{An operation on the theory is described as attaching a domain wall from the
bulk that implements the operation.
}
    \label{Top manipulation in sym tft}
\end{figure}

\paragraph{Topological  manipulation}
Now we define the physical boundary condition corresponding to our setup to be
\(\ket{T_B}\), and assume that the partition function \(Z_{T_B}[a]\) is given by
\begin{equation}
  Z_{T_B}[a] = \langle e,a | T_B \rangle .
\end{equation}
We now define the operator \(O\) acting on \(V(\Sigma)\) as follows:
\begin{equation}
  O
  = \frac{1}{\lvert H^1(\Sigma;\mathbb{Z}_2)\rvert}
    \sum_{a,b \in H^1(\Sigma;\mathbb{Z}_2)}
    (-1)^{\int a \cup b}\,
    \ket{e,a}\bra{e,b}.
\end{equation}
Using this expression, one can check that the domain wall corresponding to
\(O\) exchanges \(E(a)\) and \(M(a)\):
\begin{equation}
  O\circ E(a)\circ O^{-1} = M(a),
  \qquad
  O\circ M(a)\circ O^{-1} = E(a).\label{O on e,m}
\end{equation}
It then follows that this domain wall implements the gauging operation as follows:
\begin{equation}
    \begin{aligned}
        \langle e,A|O|T_B\rangle&=\langle e,A |\left(\frac{1}{\lvert H^1(\Sigma;\mathbb{Z}_2)\rvert}
    \sum_{a,b \in H^1(\Sigma;\mathbb{Z}_2)}
    (-1)^{\int a \cup b}\,
    \ket{e,a}\bra{e,b}\right)|T_B\rangle\\
    &=\frac{1}{|H^1(\Sigma;\ZZ_2)|^{1/2}} \sum_{b\in H^1(\Sigma;\mathbb{Z}_2)}(-1)^{\int b\cup A}Z_{T_B}[b]\\
    &=Z_{O^B(T_B)}[A].
    \end{aligned}
\end{equation}
Furthermore, by inserting the identity given in~\eqref{identity m}, we obtain
\begin{equation}
\begin{aligned}
    \langle e,A|O&=\frac{1}{|H^1(\Sigma;\ZZ_2)|} \sum_{a\in H^1(\Sigma;\mathbb{Z}_2)}(-1)^{\int a\cup A}\langle e, a |\left(\sum_{b\in H^1(\Sigma;\mathbb{Z}_2)} \ket{m,b}\bra{m,b}\right)  \\
     &=\frac{1}{|H^1(\Sigma;\ZZ_2)|} \sum_{a\in H^1(\Sigma;\mathbb{Z}_2)}(-1)^{\int a\cup (A+b)}\langle m,b| \\
        &=\langle m,A|.
\end{aligned}
\end{equation}
In this way, we can interpret the gauging operation \(O\) as a change
in the choice of the gapped boundary.
We now move on to \(S_1\), which acts as in~\eqref{op on bos theory S_1}.
Let us define \(S_1\) as follows:
\begin{equation}
  S_1
  = \frac{1}{\lvert H^1(\Sigma;\mathbb{Z}_2) \rvert^{1/2}}
    \sum_{a \in H^1(\Sigma;\mathbb{Z}_2)}
    (-1)^{\int a \cup w_1}\,
    \ket{e,a}\bra{e,a}.
\end{equation}
Using this form, one can check that the domain wall corresponding to \(S_1\)
acts trivially on \(E(a)\), while acting non-trivially on \(M(a)\) by a phase:
\begin{equation}
  S_1\circ E(a)\circ S_1^{-1} = E(a),
  \qquad
  S_1\circ M(a)\circ S_1^{-1}
  = (-1)^{\int a \cup w_1} M(a).\label{S_1 on E,M}
\end{equation}
One can easily check that 
\begin{equation}
    \langle e,A|S_1 |T_B\rangle=Z_{S_1^B(T_B)}[A]
\end{equation}
Note that although the action of \((O S_1)^4 = S_2\) on line operators is trivial,
\begin{equation}
  S_2\circ E(a)\circ S_2^{-1} = E(a),
  \qquad
  S_2\circ M(a)\circ S_2^{-1} = M(a),
\end{equation}
it acts non-trivially on \(V(\Sigma)\) by an overall phase
\(\,(-1)^{\int w_2}\).

\subsection{Fermionic theory}
Let us now move on to the fermionic case.
We have a fermionic line operator
\(
F(a) = E(a) M(a).
\)
What is non-trivial here is that
\begin{equation}
    \begin{aligned}
         F(a)^2
  &= E(a) M(a) E(a) M(a) \\
  &= (-1)^{\int a \cup a}\, E(a)^2 M(a)^2 \\
  &= (-1)^{\int a \cup a},
    \end{aligned}
\end{equation}
where the second line follows from the commutation relation
in~\eqref{commutation E and B}.
This phase is trivial in orientable cases; however, it becomes non-trivial in non-orientable cases since
we have \(a \cup a = a \cup w_1\)
as mentioned in footnote~\ref{a^2=aw_1}. 
Moreover, from ~\eqref{addition of e m} and~\eqref{commutation E and B}, we have
\begin{equation}
\begin{aligned}
  F(a+b)
  &= E(a) E(b) M(a) M(b) \\
  &= (-1)^{\int a \cup b} E(a) M(a) E(b) M(b) \\
  &= (-1)^{\int a \cup b} F(a) F(b).
\end{aligned}
\end{equation}
This precisely matches the defining property of a quadratic refinement,
\begin{equation}
  q_\eta(a+b)
  = q_\eta(a) + q_\eta(b) + 2 \int a \cup b .
\end{equation}

\paragraph{Boundary condition}
One can obtain a fermionic boundary state \(\ket{\eta}\)
satisfying
\begin{equation}
  F(a)\ket{\eta}
  = \exp\!\left(-\frac{\pi i}{2}\, q_{\eta}(a)\right)\ket{\eta},
  \qquad
  \text{for all } a \in H^1(\Sigma;\mathbb{Z}_2),
\end{equation}
where \(\eta\in\Pin^-(\Sigma)\).
Here we allow \(\Sigma\) to be non-orientable. If \(\Sigma\) is orientable, the \(\mathbb{Z}_4\)-valued quadratic refinement
reduces to a \(\mathbb{Z}_2\)-valued one and reproduces the known results, for example in~\cite{Wen:2024udn}.\footnote{For a more algebraic discussion of fermionic SymTFTs, see e.g.~\cite{Bhardwaj:2024ydc}.
}
We then easily see that
\(\ket{\eta+ b} := E(b)\ket{\eta}\)
is also an eigenstate of \(F(a)\), since
\begin{equation}
\begin{aligned}
  F(a)\ket{\eta + b}
  &= E(a) M(a) E(b)\ket{\eta} \\
  &= (-1)^{\int a \cup b}\, E(b) E(a) M(a)\ket{\eta} \\
  &= \exp\!\left(-\frac{\pi i}{2} q_{\eta}(a)\right)
     (-1)^{\int a \cup b} E(b)\ket{\eta} \\
  &= \exp\!\left(-\frac{\pi i}{2} q_{\eta + b}(a)\right)
     \ket{\eta + b}.
\end{aligned}
\end{equation}
One can further check that
\begin{equation}
  E(a)\ket{\eta + b} = \ket{\eta + a + b}.
\end{equation}
These states span the vector space associated with \(\Sigma\), namely,
\begin{equation}
  V(\Sigma)
  = \mathrm{span}_{\mathbb{C}}
    \bigl\{ \ket{\eta} \mid \eta\in \Pin^-(\Sigma) \bigr\},
\end{equation}
We fix the normalization by
\begin{align}
    \langle e,0|\eta \rangle &= 1,\\
    \langle \eta|\eta' \rangle &= |H^1(\Sigma;\mathbb{Z}_2)|^{1/2}\delta_{\eta\eta'},
\end{align}
\begin{equation}
    1=\frac{1}{|H^1(\Sigma;\mathbb{Z}_2)|^{1/2}}\sum_{\eta\in\Pin^-(\Sigma)} \ket{\eta}\bra{\eta}.
\end{equation}
Then we have
\begin{equation}
\begin{aligned}
  \langle e,a | \eta \rangle
  &= \langle e,0| E(a) M(a)| \eta\rangle \\
  &=
     \langle e,0|F(a)|\eta \rangle \\
  &= \exp\left(-\frac{\pi i}{2}q_{\eta}(a)\right)
     \langle e,0 | \eta\rangle \\
  &= \exp\left(-\frac{\pi i}{2}q_{\eta}(a)\right).
\end{aligned}\label{inner prod e f}
\end{equation}
Therefore, we have
\begin{equation}
    \langle \eta | e,a\rangle= \exp\left(\frac{\pi i}{2}q_{\eta}(a)\right).
\end{equation}
This implies that we can write
\begin{equation}
    \ket{e,a}=\frac{1}{|H^1(\Sigma;\mathbb{Z}_2)|^{1/2}}\sum_{\eta\in\Pin^-(\Sigma)}\exp\left(\frac{\pi i}{2}q_{\eta}(a)\right)\ket{\eta}.
\end{equation}
One can also compute
\begin{equation}
    \begin{aligned}
        \langle m,a|\eta\rangle&=\frac{1}{|H^1(\Sigma;\bZ_2)|^{1/2}}\sum_{c\in H^1(\Sigma;\bZ_2)}\langle m,a|e,c\rangle \langle e,c|\eta\rangle\\
        &=\frac{1}{|H^1(\Sigma;\bZ_2)|^{1/2}}\sum_{c\in H^1(\Sigma;\bZ_2)}\exp\left(-\frac{\pi i}{2}q_{\eta+a}(c)\right)\\
        &=\exp\left(-\frac{\pi i}{4}\mathrm{ABK}(\eta+a)\right).
    \end{aligned}
    \label{eq:maABK}
\end{equation}
This implies 
\begin{equation}
    \ket{m,a}=\frac{1}{|H^1(\Sigma;\mathbb{Z}_2)|^{1/2}}\sum_{\eta\in\Pin^-(\Sigma)}\exp\left(\frac{\pi i}{4}\mathrm{ABK}(\eta+a)\right)\ket{\eta}.
\end{equation}

\paragraph{Topological manipulation}
Then we can interpret the fermionization operation \(\mathrm{Fer}\) as a change
in the choice of the gapped boundary:\footnote{Since $\braket{e,a|m,0}=1$, $\ket{m,0}$ is the physical boundary condition corresponding to the bosonic trivial theory $\mathrm{Tri}_B$. We can interpret that $\braket{\eta|m,0}=Z_\mathrm{ABK}[\eta]$ reflects $\Fer(\mathrm{Tri}_B)=\ABK$.}
\begin{equation}
    \begin{aligned}
      Z_{\Fer(T_B)}[\eta]&=\frac{1}{|H^1(\Sigma;\bZ_2)|^{1/2}}\sum_{a\in H^1(\Sigma;\bZ_2)}\exp\left(\frac{\pi i}{2}q_{\eta}(a)\right)Z_{T_B}[a]\\
       &=\frac{1}{|H^1(\Sigma;\bZ_2)|^{1/2}}\sum_{a\in H^1(\Sigma;\bZ_2)}\langle\eta|e,a\rangle\langle e,a|T_B\rangle\\
       &=\langle\eta|T_B\rangle.
    \end{aligned}
\end{equation}
Then let us compute \(\bra{\eta} O\ket{T_B}\) as follows:
\begin{equation}
\begin{aligned}
  \bra{\eta} O \ket{T_B}
  &=
  \bra{\eta}
  \left(
    \frac{1}{\lvert H^1(\Sigma;\mathbb{Z}_2)\rvert}
    \sum_{a,b \in H^1(\Sigma;\mathbb{Z}_2)}
    (-1)^{\int a \cup b}\,
    \ket{e,a}\bra{e,b}
  \right)
  \ket{T_B} \\
  &=
  \frac{1}{\lvert H^1(\Sigma;\mathbb{Z}_2)\rvert}
  \sum_{a,b \in H^1(\Sigma;\mathbb{Z}_2)}
  \exp\!\left(\frac{\pi i}{2}\,
    q_{\eta + b}(a)\right)
  Z_{T_B}[b] \\
  &=
  \frac{1}{\lvert H^1(\Sigma;\mathbb{Z}_2)\rvert^{1/2}}
  \sum_{b \in H^1(\Sigma;\mathbb{Z}_2)}
  Z_{\mathrm{ABK}}[\eta + b]\,
  Z_{T_B}[b].
\end{aligned}
\end{equation}
Here, using~\eqref{abk and quad}, 
\begin{equation}
\begin{aligned}
  Z_{\mathrm{ABK}}[\eta + b]
  &=
  Z_{\mathrm{ABK}}[\eta]\,
  \exp\!\left(\frac{\pi i}{2}\,
    q_{\eta}(b)\right)
  (-1)^{\int b \cup b} \\
  &=
  Z_{\mathrm{ABK}}[\eta]\,
  \exp\!\left(\frac{\pi i}{2}\,
    q_{\eta+ w_1}(b)\right).
\end{aligned}
\end{equation}
Therefore, we obtain
\begin{equation}
\begin{aligned}
  \bra{\eta} O \ket{T_B}
  &=
  \frac{1}{\lvert H^1(\Sigma;\mathbb{Z}_2)\rvert^{1/2}}
  \sum_{b \in H^1(\Sigma;\mathbb{Z}_2)}
  Z_{\mathrm{ABK}}[\eta]\,
  \exp\!\left(\frac{\pi i}{2}\,
    q_{\eta + w_1}(b)\right)
  Z_{T_B}[b] \\
  &=
  Z_{\mathrm{ABK}}[\eta]\,
  Z_{\Fer(T_B)}[\eta + w_1]\\
  &=Z_{O^F\circ \Fer(T_B)}[\eta].
\end{aligned}
\end{equation}
Furthermore, by computing $\bra{\eta} S_1 \ket{T_B}$, we find that
\begin{equation}
  \begin{aligned}
        \bra{\eta} S_1 \ket{T_B}
    &= Z_{\Fer(T_B)}[\eta + w_1]\\
    &=Z_{S_1^F\circ \Fer (T_B)}[\eta]\, .
  \end{aligned}
\end{equation}
Accordingly, we find that \(O\) acts trivially, while \(S_1\) acts on \(F(a)\)
by a phase:
\begin{equation}
  O \circ F(a) \circ O^{-1}
  = F(a),\qquad
  S_1 \circ F(a) \circ S_1^{-1}
  =
  (-1)^{\int a \cup a}\, F(a),
\end{equation}
where we have used~\eqref{O on e,m} and~\eqref{S_1 on E,M}.
As in the bosonic case, note that although the action of \((O S_1)^4 = S_2\) on line operators is trivial,
\begin{equation}
  S_2\circ F(a)\circ S_2^{-1} = F(a),
\end{equation}
it acts non-trivially on \(V(\Sigma)\) by an overall phase
\(\,(-1)^{\int w_2}\):
\begin{equation}
  \begin{aligned}
        \bra{\eta}S_2\ket{T_B}&=Z_{\Fer(T_B)}[\eta](-1)^{\int w_2}\\
    &=Z_{S_2^F\circ \Fer (T_B)}[\eta].
  \end{aligned}
\end{equation}

%% file: 4.tex
\section{Dualities of sectors}
In this section, we place theories on a circle $S^1$. For a bosonic theory with a $\mathbb{Z}_2$ symmetry, the Hilbert space on $S^1$ is $\mathbb{Z}_2$ untwisted or twisted. We can decompose the untwisted and twisted states according to the eigenvalues of the $\ZZ_2$ symmetry and the parity transformation $P$. Since $P^2=1$ in a bosonic theory $T_B$, the eigenvalues of $P$ are $\pm1$.\footnote{Note that in a $\Pin^-$ theory, the eigenvalues of $P$ are $+1,-1,+i$ and $-i$ since $P^2=(-1)^F$.} The decomposition is summarized in Table~\ref{table:decomposition}, where $\mathsf{S}_\pm,\mathsf{T}_\pm,\mathsf{U}_\pm,$ and $\mathsf{V}_\pm$ are symbols for respective sectors. These symbols are defined by extending the notation in \cite{Hsieh:2020uwb} to include the eigenvalues of $P$. A summary of the results is given in Table~\ref{table:Summary}.

\newcolumntype{P}[1]{>{\centering\arraybackslash}p{#1}}
\newcommand{\sSp}{\cellcolor{cyan!40}$\mathsf{S}_+$}
\newcommand{\sSm}{\cellcolor{cyan!20}$\mathsf{S}_-$}
\newcommand{\sTp}{\cellcolor{YellowGreen!40}$\mathsf{T}_+$}
\newcommand{\sTm}{\cellcolor{YellowGreen!20}$\mathsf{T}_-$}
\newcommand{\sUp}{\cellcolor{red!40}$\mathsf{U}_+$}
\newcommand{\sUm}{\cellcolor{red!20}$\mathsf{U}_-$}
\newcommand{\sVp}{\cellcolor{yellow!40}$\mathsf{V}_+$}
\newcommand{\sVm}{\cellcolor{yellow!10}$\mathsf{V}_-$}

\begin{table}[h]
\centering
\begin{tabular}{c|c|c|c}
    \multicolumn{2}{c|}{$T_B$} & untwisted & twisted \\ \hline
    \multirow{2}{2em}{even} & $P=+1$ & \sSp & \sUp \\
    & $P=-1$ & \sSm & \sUm \\ \hline
    \multirow{2}{2em}{odd} & $P=+1$ & \sTp & \sVp \\
    & $P=-1$ & \sTm & \sVm
\end{tabular}
\caption{Sectors of the bosonic theories $T_B$ on the circle.}
\label{table:decomposition}
\end{table}

\begin{table}[h]
\centering
\footnotesize
\begin{minipage}[c]{0.32\columnwidth}
    \centering
    \begin{tabular}{c|c|c|c}
        \multicolumn{2}{c|}{$T_B$} & $\mathcal{H}$ & $\mathcal{H}^g$  \\ \hline
        \multirow{2}{2em}{even} & $P=+1$ & \sSp & \sUp \\
                                & $P=-1$ & \sSm & \sUm \\ \hline
        \multirow{2}{2em}{odd}  & $P=+1$ & \sTp & \sVp \\
                                & $P=-1$ & \sTm & \sVm
    \end{tabular}
\end{minipage}
\begin{minipage}[c]{0.32\columnwidth}
    \centering
    \begin{tabular}{c|c|c|c}
        \multicolumn{2}{c|}{$O^B(T_B)$} & $\mathcal{H}$ & $\mathcal{H}^g$ \\ \hline
        \multirow{2}{2em}{even} & $P=+1$ & \sSp & \sTp \\
                                & $P=-1$ & \sSm & \sTm \\ \hline
        \multirow{2}{2em}{odd}  & $P=+1$ & \sUp & \sVm \\
                                & $P=-1$ & \sUm & \sVp
    \end{tabular}
\end{minipage}
\begin{minipage}[c]{0.32\columnwidth}
    \centering
    \begin{tabular}{c|c|c|c}
        \multicolumn{2}{c|}{$S_1^B(T_B)$} & $\mathcal{H}$ & $\mathcal{H}^g$ \\ \hline
        \multirow{2}{2em}{even} & $P=+1$ & \sSp & \sUm \\
                                & $P=-1$ & \sSm & \sUp \\ \hline
        \multirow{2}{2em}{odd}  & $P=+1$ & \sTp & \sVm \\
                                & $P=-1$ & \sTm & \sVp
    \end{tabular}
\end{minipage}
\\[3mm]
\begin{minipage}[c]{0.32\columnwidth}
    \centering
    \begin{tabular}{c|P{6mm}|P{6mm}}
        $T_F$ & NS & R \\ \hline
        $P=+1$ & \sSp & \sTp \\
        $P=-1$ & \sSm & \sTm \\
        $P=+i$ & \sVp & \sUm \\
        $P=-i$ & \sVm & \sUp
    \end{tabular}
\end{minipage}
\begin{minipage}[c]{0.32\columnwidth}
    \centering
    \begin{tabular}{c|P{6mm}|P{6mm}}
        $O^F(T_F)$ & NS & R \\ \hline
        $P=+1$ & \sSp & \sUp \\
        $P=-1$ & \sSm & \sUm \\
        $P=+i$ & \sVm & \sTm \\
        $P=-i$ & \sVp & \sTp
    \end{tabular}
\end{minipage}
\begin{minipage}[c]{0.32\columnwidth}
    \centering
    \begin{tabular}{c|P{6mm}|P{6mm}}
        $S_1^F(T_F)$ & NS & R \\ \hline
        $P=+1$ & \sSp & \sTp \\
        $P=-1$ & \sSm & \sTm \\
        $P=+i$ & \sVm & \sUp \\
        $P=-i$ & \sVp & \sUm
    \end{tabular}
\end{minipage}
\caption{Summary of the dualities among sectors.}
\label{table:Summary}
\end{table}

\subsection{Bosonic theory}
When no $\ZZ_2$ gauge line is inserted, the torus partition function is the trace:
\begin{equation}
    Z_{T_B}[T^2;0]=\Tr^{T_B}_\mathcal{H}(e^{-lH}),
\end{equation}
where $\mathcal{H}$ is the untwisted Hilbert space, $H$ is the Hamiltonian, and $l$ is the length of the torus in the temporal direction.

\begin{figure}[h]
    \centering
    \begin{tikzpicture}[auto, scale=1]
        \draw[dashed] (1.5,0) arc[start angle=0, end angle=180, x radius=1.5, y radius=0.6];
        \draw (1.5,0) arc[start angle=0, end angle=-180, x radius=1.5, y radius=0.6];
        \draw (0,3) circle[x radius=1.5, y radius=0.6];
        \draw (1.5,0) -- (1.5,3);
        \draw (-1.5,0) -- (-1.5,3);

        \draw[red,thick,dashed] (1.5,1.5) arc[start angle=0, end angle=180, x radius=1.5, y radius=0.6];
        \draw[red,thick] (1.5,1.5) arc[start angle=0, end angle=-180, x radius=1.5, y radius=0.6] node[left] {$\textcolor{red}{x}$};
        \draw[cyan,thick] (0,-0.6) -- (0,2.4) node[above] {$\textcolor{cyan}{y}$};

        \draw[->] (2.5,0) --node[midway, right] {time} (2.5,3);
    \end{tikzpicture}
    \caption{The torus $T^2$ is represented as a cylinder with its ends identified, with time running upward. The Poincaré duals of the spatial and temporal cycles define $x,y \in H^1(T^2;\mathbb{Z}_2)$. For the Klein bottle, an orientation-reversing defect is inserted along the spatial direction, which affects the computation. See~\eqref{int on K}.}
    \label{fig:torus_lines}
\end{figure}
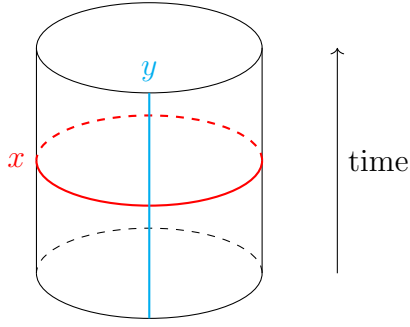

As in Figure~\ref{fig:torus_lines}, we denote $x,y\in H^1(T^2;\ZZ_2)$ the Poincaré duals of the spatial and temporal loops, respectively. Since $y$ is the $\ZZ_2$ twist, the torus partition function with $y$ inserted is
\begin{equation}
    Z_{T_B}[T^2;y]=\Tr^{T_B}_{\mathcal{H}^g}(e^{-lH}),
\end{equation}
where $\mathcal{H}^g$ is the twisted Hilbert space. Inserting $x$ corresponds to inserting the $\ZZ_2$ symmetry operator $g$ in the trace:
\begin{equation}
    Z_{T_B}[T^2;x]=\Tr^{T_B}_\mathcal{H}(g\, e^{-lH}).
\end{equation}

If we insert the parity transformation $P$ in the spatial direction,\footnote{Equivalently, this can be described as inserting the Poincaré dual of \(w_1\)
along the spatial direction.
Crossing this defect flips the orientation.
See e.g.~\cite{Thorngren:2014pza}.
} the partition function on torus $T^2$  becomes one on the Klein bottle $K$. As in the torus case, we denote $x,y\in H^1(K;\ZZ_2)$ the Poincaré duals of the loops in the spatial and temporal directions, respectively. The partition functions on $K$ are calculated by inserting $P$:
\begin{align}
    Z_{T_B}[K;0]&=\Tr^{T_B}_{\mathcal{H}}(P e^{-lH}),\\
    Z_{T_B}[K;x]&=\Tr^{T_B}_{\mathcal{H}}(g P e^{-lH}),\\
    Z_{T_B}[K;y]&=\Tr^{T_B}_{\mathcal{H}^g}(P e^{-lH}),\\
    Z_{T_B}[K;x+y]&=\Tr^{T_B}_{\mathcal{H}^g}(g P e^{-lH}).
\end{align}
The traces over the sectors can be expressed as the sums of the partition functions. For example, the trace over the $\ZZ_2$-even and $P=+1$ sector of the untwisted Hilbert space $\mathcal{H}$ is
\begin{equation}
    \Tr^{T_B}_\mathcal{H}\left(\frac{1+g}{2}\frac{1+P}{2} e^{-l H}\right)=\frac{1}{4}\left(Z_{T_B}[T^2;0]+Z_{T_B}[T^2;x]+Z_{T_B}[K;0]+Z_{T_B}[K;x]\right).
\end{equation}
Another example is the trace over the $\ZZ_2$-odd and $P=-1$ sector of the twisted Hilbert space $\mathcal{H}^g$:
\begin{equation}
    \Tr^{T_B}_{\mathcal{H}^g}\left(\frac{1-g}{2}\frac{1-P}{2} e^{-l H}\right)=\frac{1}{4}\left(Z_{T_B}[T^2;y]-Z_{T_B}[T^2;x+y]-Z_{T_B}[K;y]+Z_{T_B}[K;x+y]\right).
\end{equation}

The cup product of $H^\bullet(T^2;\ZZ_2)$ can be calculated using
\begin{equation}
    x\cup x=0,\quad y\cup y=0,\quad x\cup y=1\quad \text{on}\ T^2.
\end{equation}
The cup product of $H^\bullet(K;\ZZ_2)$ is not the same:
\begin{equation}
    x\cup x=0,\quad y\cup y=1,\quad x\cup y=1\quad \text{on}\ K.\label{int on K}
\end{equation}
This difference reflects that the first Stiefel--Whitney class is $w_1=0$ on $T^2$ while $w_1=x$ on $K$. Generally, $A \in H^1(\Sigma;\ZZ_2)$ satisfies $A\cup A=A\cup w_1,$ so a non-vanishing $w_1$ changes the calculation. The second Stiefel-Whitney class $w_2$ on two-dimensional closed manifolds satisfies
\begin{equation}
    w_2=w_1\cup w_1.
\end{equation}
Since $w_2=0$ on $T^2$ and $K$, the equation $x\cup x=0$ is unchanged.

We can calculate the action of gauging $O^B$ on the torus partition functions:
\begin{equation}
    \begin{pmatrix}
        Z_{O^B(T_B)}[T^2;0] \\ Z_{O^B(T_B)}[T^2;x] \\ Z_{O^B(T_B)}[T^2;y] \\ Z_{O^B(T_B)}[T^2;x+y]
    \end{pmatrix}
    =\frac{1}{2}
    \begin{pmatrix}
        1 & 1 & 1 & 1\\
        1 & 1 & -1 & -1\\
        1 & -1 & 1 & -1\\
        1 & -1 & -1 & 1
    \end{pmatrix}
    \begin{pmatrix}
        Z_{T_B}[T^2;0] \\ Z_{T_B}[T^2;x] \\ Z_{T_B}[T^2;y] \\ Z_{T_B}[T^2;x+y]
    \end{pmatrix}\\,
\end{equation}
and the Klein bottle partition functions:
\begin{equation}
    \begin{pmatrix}
        Z_{O^B(T_B)}[K;0] \\ Z_{O^B(T_B)}[K;x] \\ Z_{O^B(T_B)}[K;y] \\ Z_{O^B(T_B)}[K;x+y]
    \end{pmatrix}
    =\frac{1}{2}
    \begin{pmatrix}
        1 & 1 & 1 & 1\\
        1 & 1 & -1 & -1\\
        1 & -1 & -1 & 1\\
        1 & -1 & 1 & -1
    \end{pmatrix}
    \begin{pmatrix}
        Z_{T_B}[K;0] \\ Z_{T_B}[K;x] \\ Z_{T_B}[K;y] \\ Z_{T_B}[K;x+y]
    \end{pmatrix}.
\end{equation}
The difference between $T^2$ and $K$ arises when $A=y,x+y$. We can then deduce the relations of the traces. For example:
\begin{equation}
    \Tr^{O^B(T_B)}_\mathcal{H}\left(\frac{1+g}{2}\frac{1+P}{2} e^{-l H}\right)=\Tr^{T_B}_\mathcal{H}\left(\frac{1+g}{2}\frac{1+P}{2} e^{-l H}\right).
\end{equation}
This means that the $\ZZ_2$-even and $P=+1$ sectors of the untwisted Hilbert space are the same in $O^B(T_B)$ and $T_B$, which we denote $\mathsf{S}_+$.

The action of $S_1^B$ is easier. Since $w_1=0$ on $T^2$, the torus partition functions are the same:
\begin{equation}
    Z_{S_1^B(T_B)}[T^2;A]=Z_{T_B}[T^2;A].
\end{equation}
The Klein bottle partition functions flip the sign when $A=y,x+y$:
\begin{equation}
    \begin{pmatrix}
        Z_{S_1^B(T_B)}[T^2;0] \\ Z_{S_1^B(T_B)}[T^2;x] \\ Z_{S_1^B(T_B)}[T^2;y] \\ Z_{S_1^B(T_B)}[T^2;x+y]
    \end{pmatrix}
    =
    \begin{pmatrix}
        1 & 0 & 0 & 0\\
        0 & 1 & 0 & 0\\
        0 & 0 & -1 & 0\\
        0 & 0 & 0 & -1
    \end{pmatrix}
    \begin{pmatrix}
        Z_{T_B}[T^2;0] \\ Z_{T_B}[T^2;x] \\ Z_{T_B}[T^2;y] \\ Z_{T_B}[T^2;x+y]
    \end{pmatrix}.
\end{equation}

\begin{table}[h]
\centering
\begin{minipage}[c]{0.45\columnwidth}
    \centering
    \begin{tabular}{c|c|c|c}
        \multicolumn{2}{c|}{$O^B(T_B)$} & untwisted & twisted \\ \hline
        \multirow{2}{2em}{even} & $P=+1$ & \sSp & \sTp \\
        & $P=-1$ & \sSm & \sTm \\ \hline
        \multirow{2}{2em}{odd} & $P=+1$ & \sUp & \sVm \\
        & $P=-1$ & \sUm & \sVp
    \end{tabular}
\end{minipage}
\begin{minipage}[c]{0.45\columnwidth}
    \centering
    \begin{tabular}{c|c|c|c}
        \multicolumn{2}{c|}{$S_1^B(T_B)$} & untwisted & twisted \\ \hline
        \multirow{2}{2em}{even} & $P=+1$ & \sSp & \sUm \\
        & $P=-1$ & \sSm & \sUp \\ \hline
        \multirow{2}{2em}{odd} & $P=+1$ & \sTp &\sVm\\
        & $P=-1$ & \sTm & \sVp
    \end{tabular}
\end{minipage}
\caption{States of the bosonic theories $O^B(T_B)$ and $S_1^B(T_B)$.}
\label{table:states_bosonic}
\end{table}

The relations of the sectors are summarized in Table~\ref{table:states_bosonic}. $O^B$ exchanges $\mathsf{T}_\pm$ and $\mathsf{U}_\pm$, and exchanges $\mathsf{V}_+$ and $\mathsf{V}_-$. $S_1^B$ exchanges $\mathsf{U}_+$ and $\mathsf{U}_-$, and exchanges $\mathsf{V}_+$ and $\mathsf{V}_-$.

Since $w_2=0$ on both $T^2$ and $K$, the action of $S_2^B$ is trivial:
\begin{equation}
    Z_{S_2^B(T_B)}[T^2;A]=Z_{T_B}[T^2;A],\quad Z_{S_2^B(T_B)}[K;A]=Z_{T_B}[K;A],
\end{equation}
so $S_2^B$ does not affect the Hilbert space on $S^1$.

\subsection{Fermionic theory}
We choose a reference $\Pin^-$ structure $\eta_0$ on $T^2$ whose quadratic refinement is
\begin{equation}
    q_{\eta_0}(0)=0,\quad q_{\eta_0}(x)=0,\quad q_{\eta_0}(y)=0,\quad q_{\eta_0}(x+y)=2.
\end{equation}
This corresponds to the NSNS boundary condition, in which a fermion field is anti-periodic in spatial and temporal direction. The relation $q_{\eta+a}(b)=q_\eta(b)+2\int a\cup b$ and the formulae of the cup products give
\begin{align}
    q_{\eta_0+x}(0)&=0,&q_{\eta_0+x}(x)&=0,&q_{\eta_0+x}(y)&=2,&q_{\eta_0+x}(x+y)&=0,\\
    q_{\eta_0+y}(0)&=0,&q_{\eta_0+y}(x)&=2,&q_{\eta_0+y}(y)&=0,&q_{\eta_0+y}(x+y)&=0,\\
    q_{\eta_0+x+y}(0)&=0,&q_{\eta_0+x+y}(x)&=2,&q_{\eta_0+x+y}(y)&=2,&q_{\eta_0+x+y}(x+y)&=2.
\end{align}
Since a torus $T^2$ is orientable, the value of the quadratic refinement of a $\Pin^-$ structure is twice the value of the quadratic refinement of the corresponding spin structure.

For $K$, we choose a reference $\Pin^-$ structure $\eta'_0$ whose quadratic refinement is
\begin{equation}
    q_{\eta'_0}(0)=0,\quad q_{\eta'_0}(x)=0,\quad q_{\eta'_0}(y)=1,\quad q_{\eta'_0}(x+y)=3.
\end{equation}
This corresponds to the spatial NS boundary condition. The quadratic refinements of the other $\Pin^-$ structures are
\begin{align}
    q_{\eta'_0+x}(0)&=0,&q_{\eta'_0+x}(x)&=0,&q_{\eta'_0+x}(y)&=3,&q_{\eta'_0+x}(x+y)&=1,\\
    q_{\eta'_0+y}(0)&=0,&q_{\eta'_0+y}(x)&=2,&q_{\eta'_0+y}(y)&=3,&q_{\eta'_0+y}(x+y)&=3,\\
    q_{\eta'_0+x+y}(0)&=0,&q_{\eta'_0+x+y}(x)&=2,&q_{\eta'_0+x+y}(y)&=1,&q_{\eta'_0+x+y}(x+y)&=1.
\end{align}

In a bosonic theory, the parity transformation $P$ is an involution: $P^2=1$. However, in a $\Pin^-$ theory, $P^2$ is the fermion parity $(-1)^F$. Since $P^4=1$, the eigenvalues of $P$ are $+1,-1,+i,$ and $-i$. We can decompose the NS Hilbert space $\mathcal{H}_\text{NS}$ and the R Hilbert space $\mathcal{H}_\text{R}$ into four eigenspaces of $P$.

When we add $x$ to a $\Pin^-$ structure, $(-1)^F=P^2$ is inserted in the trace:\footnote{If one takes partition functions as the starting point, choosing a reference $\Pin^-$ structure $\eta'_0$ on the Klein bottle $K$ is equivalent to defining a parity transformation $P$. One may instead choose $\eta'_0+x$ as the reference, which corresponds to redefining the parity as $P^3=P(-1)^F$. Since $w_1=x$ on $K$, shifting $\eta'_0$ to $\eta'_0+x$ is equivalent to the action of $S_1^F$. We investigate the connection between the operations and redefinition of parity in the next section in more detail.}
\begin{align}
    Z_{T_F}[T^2;\eta_0]&=\Tr_{\mathcal{H}_\text{NS}}(e^{-lH}),\\
    Z_{T_F}[K;\eta'_0]&=\Tr_{\mathcal{H}_\text{NS}}(Pe^{-lH}),\\
    Z_{T_F}[T^2;\eta_0+x]&=\Tr_{\mathcal{H}_\text{NS}}(P^2e^{-lH}),\\
    Z_{T_F}[K;\eta'_0+x]&=\Tr_{\mathcal{H}_\text{NS}}(P^3e^{-lH}).
\end{align}
Adding $y$ changes the Hilbert space from $\mathcal{H}_\text{NS}$ to $\mathcal{H}_\text{R}$. Thus, as in the bosonic case, the trace over the sectors can be expressed as a sum of the partition functions. For example, the trace over the $P=+1$ sector in $\mathcal{H}_\text{NS}$ is
\begin{equation}
    \begin{aligned}
        &\Tr^{T_F}_{\mathcal{H}_\text{NS}}\left(\frac{1+P+P^2+P^3}{4} e^{-l H}\right)\\
        &=\frac{1}{4}(Z_{T_F}[T^2;\eta_0]+Z_{T_F}[K;\eta'_0]+Z_{T_F}[T^2;\eta_0+x]+Z_{T_F}[K;\eta'_0+x]).
    \end{aligned}
\end{equation}
Another example is the trace over the $P=+i$ sector in $\mathcal{H}_\text{R}$:
\begin{equation}
    \begin{aligned}
        &\Tr^{T_F}_{\mathcal{H}_\text{R}}\left(\frac{1-iP-P^2+iP^3}{4} e^{-l H}\right)\\
        &=\frac{1}{4}(Z_{T_F}[T^2;\eta_0+y]-iZ_{T_F}[K;\eta'_0+y]-Z_{T_F}[T^2;\eta_0+x+y]+iZ_{T_F}[K;\eta'_0+x+y]).
    \end{aligned}
\end{equation}

Recall that the fermionization formula is
\begin{equation}
    Z_{\Fer(T_B)}[\Sigma;\eta]=\frac{1}{|H^1(\Sigma;\ZZ_2)|^{1/2}}\sum_{a\in H^1(\Sigma;\ZZ_2)}Z_{T_B} [\Sigma;a]\exp\left(\frac{\pi i}{2}q_\eta(a)\right),
\end{equation}
so the torus partition functions are
\begin{equation}
    \begin{pmatrix}
        Z_{\Fer(T_B)}[T^2;\eta_0] \\ Z_{\Fer(T_B)}[T^2;\eta_0+x] \\ Z_{\Fer(T_B)}[T^2;\eta_0+y] \\ Z_{\Fer(T_B)}[T^2;\eta_0+x+y]
    \end{pmatrix}
    =\frac{1}{2}
    \begin{pmatrix}
        1 & 1 & 1 & -1\\
        1 & 1 & -1 & 1\\
        1 & -1 & 1 & 1\\
        1 & -1 & -1 & -1
    \end{pmatrix}
    \begin{pmatrix}
        Z_{T_B}[T^2;0] \\ Z_{T_B}[T^2;x] \\ Z_{T_B}[T^2;y] \\ Z_{T_B}[T^2;x+y]
    \end{pmatrix},
\end{equation}
and the Klein bottle partition functions are
\begin{equation}
    \begin{pmatrix}
        Z_{\Fer(T_B)}[K;\eta'_0] \\ Z_{\Fer(T_B)}[K;\eta'_0+x] \\ Z_{\Fer(T_B)}[K;\eta'_0+y] \\ Z_{\Fer(T_B)}[K;\eta'_0+x+y]
    \end{pmatrix}
    =\frac{1}{2}
    \begin{pmatrix}
        1 & 1 & i & -i\\
        1 & 1 & -i & i\\
        1 & -1 & -i & -i\\
        1 & -1 & i & i
    \end{pmatrix}
    \begin{pmatrix}
        Z_{T_B}[K;0] \\ Z_{T_B}[K;x] \\ Z_{T_B}[K;y] \\ Z_{T_B}[K;x+y]
    \end{pmatrix}.
\end{equation}

Then, we can derive the correspondence between sectors before and after fermionization. The result is summarized in Table~\ref{table:states_fermionic}.

\begin{table}[h]
\centering
\begin{minipage}[c]{0.32\columnwidth}
    \centering
    \begin{tabular}{c|c|c}
        $T_F$ & NS & R \\ \hline
        $P=+1$ & \sSp & \sTp \\
        $P=-1$ & \sSm & \sTm \\
        $P=+i$ & \sVp & \sUm \\
        $P=-i$ & \sVm &  \sUp
    \end{tabular}
\end{minipage}
\begin{minipage}[c]{0.32\columnwidth}
    \centering
    \begin{tabular}{c|c|c}
        $O^F(T_F)$ & NS & R \\ \hline
        $P=+1$ & \sSp &  \sUp \\
        $P=-1$ &\sSm & \sUm  \\
        $P=+i$ & \sVm &\sTm \\
        $P=-i$ & \sVp & \sTp
    \end{tabular}
\end{minipage}
\begin{minipage}[c]{0.32\columnwidth}
    \centering
    \begin{tabular}{c|c|c}
        $S_1^F(T_F)$ & NS & R \\ \hline
        $P=+1$ & \sSp & \sTp \\
        $P=-1$ & \sSm & \sTm \\
        $P=+i$ & \sVm & \sUp \\
        $P=-i$ & \sVp & \sUm 
    \end{tabular}
\end{minipage}
\caption{States of the fermionic theories $T_F=\Fer(T_B),O^F(T_F)$ and $S^F_1(T_F)$.}
\label{table:states_fermionic}
\end{table}

We can derive the actions of $O^F$ and $S_1^F$ without fermionization. The partition functions of $S_1^F(T_F)$ are
\begin{equation}
    Z_{S_1^F(T_F)}[T^2;\eta]=Z_{T_F}[T^2;\eta],\quad Z_{S_1^F(T_F)}[K;\eta]=Z_{T_F}[K;\eta+x],
\end{equation}
so the trace over $P=\lambda$ sector in $\mathcal{H}_\mathrm{NS}$ of $S_1(T_F)$ is
\begin{equation}
    \begin{aligned}
        &\mathrm{Tr}^{S_1^F(T_F)}_{\mathcal{H}_\mathrm{NS}}\left(\frac{1+\lambda^{-1} P+\lambda^{-2} P^2+\lambda^{-3} P^3}{4}e^{-lH}\right)\\
        &=\mathrm{Tr}^{T_F}_{\mathcal{H}_\mathrm{NS}}\left(\frac{1+\lambda^{-1} P^3+\lambda^{-2} P^2+\lambda^{-3} P}{4}e^{-lH}\right)\\
        &=\mathrm{Tr}^{T_F}_{\mathcal{H}_\mathrm{NS}}\left(\frac{1+\lambda P+\lambda^2 P^2+\lambda^3 P^3}{4}e^{-lH}\right).
    \end{aligned}
\end{equation}
The same holds in $\mathcal{H}_\mathrm{R}$. Thus, $S_1^F$ exchanges the $P=\lambda$ sector with the $P=\lambda^{-1}=\lambda^*$ sector.

$O^F S_1^F$ is stacking the ABK theory, whose partition functions are
\begin{equation}
    \begin{aligned}
        Z_\ABK[T^2;\eta_0]=Z_\ABK[T^2;\eta_0+x]=Z_\ABK[T^2;\eta_0+y]=1,\ Z_\ABK[T^2;\eta_0+x+y]=-1,\\
        Z_\ABK[K;\eta'_0]=Z_\ABK[K;\eta'_0+x]=1,\ Z_\ABK[K;\eta'_0+y]=-i,\ Z_\ABK[K;\eta'_0+x+y]=i.
    \end{aligned}
\end{equation}
$O^F S_1^F$ does not affect $\mathcal{H}_\mathrm{NS}$. The trace over $P=\lambda$ sector in $\mathcal{H}_\mathrm{R}$ of $S_1(T_F)$ is
\begin{equation}
    \begin{aligned}
        &\mathrm{Tr}^{O^FS_1^F(T_F)}_{\mathcal{H}_\mathrm{R}}\left(\frac{1+\lambda^{-1} P+\lambda^{-2} P^2+\lambda^{-3} P^3}{4}e^{-lH}\right)\\
        &=\mathrm{Tr}^{T_F}_{\mathcal{H}_\mathrm{R}}\left(\frac{1-i\lambda^{-1} P-\lambda^{-2} P^2+i\lambda^{-3} P^3}{4}e^{-lH}\right)\\
        &=\mathrm{Tr}^{T_F}_{\mathcal{H}_\mathrm{R}}\left(\frac{1+(i\lambda)^{-1} P+(i\lambda)^2 P^2+(i\lambda)^{-3} P^3}{4}e^{-lH}\right).
    \end{aligned}
\end{equation}
Thus, the $P=\lambda$ sector in $O^FS_1^F(T_F)$ is the $P=i\lambda$ sector in $T_F$. $(O^FS_1^F)^4$ restores all sectors as in Table~\ref{table:states_RR}. This result reflects that the actions of $S_2^F=(O^FS_1^F)^4$ on $T^2$ and $K$ are trivial. If we focus only on the sectors of the $S^1$ Hilbert space, the group structure degenerates into
\begin{equation}
    D_4=D_8/\langle S_2^F\rangle=\langle O^F,S_1^F\mid (O^F)^2=(S_1^F)^2=(O^FS_1^F)^4=1\rangle,
\end{equation}
which is the dihedral group of order 8.

\begin{table}[H]
\centering
\begin{tabular}{c|P{24mm}|P{24mm}|P{24mm}|P{24mm}|P{24mm}}
    & $T_F$ & $O^F S_1^F(T_F)$ & $(O^F S_1^F)^2(T_F)$ & $(O^F S_1^F)^3(T_F)$ & $(O^F S_1^F)^4(T_F)$ \\ \hline
    $P=+1$ & \sTp & \sUm & \sTm & \sUp & \sTp \\
    $P=+i$ & \sUm & \sTm & \sUp & \sTp & \sUm \\
    $P=-1$ & \sTm & \sUp & \sTp & \sUm & \sTm \\
    $P=-i$ & \sUp & \sTp & \sUm & \sTm & \sUp
\end{tabular}
\caption{States of the R sector of the theories $T_F,O^F S_1^F(T_F),(O^F S_1^F)^2(T_F),(O^F S_1^F)^3(T_F),$ and $(O^F S_1^F)^4(T_F)$.}
\label{table:states_RR}
\end{table}

%% file: 5.tex
\section{Example: Majorana CFT}
In this section, we examine a simple but non-trivial example: the Ising CFT and the free massless Majorana fermion CFT. The Lagrangian density of the Majorana CFT is
\begin{equation}
    \mathcal{L}=i\psi_L(\partial_t-\partial_x)\psi_L+i\psi_R(\partial_t+\partial_x)\psi_R,
\end{equation}
where $\psi_L,\psi_R$ are the left-moving and right-moving Majorana fermions, respectively.

This Lagrangian is invariant under the fermion parity $(-1)^F$, the chiral fermion parity $(-1)^{F_L}$ and the naive parity transformation $\tilde{P}$:\footnote{For an invertible symmetry operator $U$ and a field $\psi$, we denote the action of $U$ on $\psi$ as $U:\psi(t,x)\to U\psi(t,x)U^{-1}$.}
\begin{align}
    (-1)^F&:\psi_L\to-\psi_L,\quad\psi_R\to-\psi_R,\\
    (-1)^{F_L}&:\psi_L\to-\psi_L,\quad\psi_R\to\psi_R,\\
    \tilde{P}&:\psi_L(t,x)\to\psi_R(t,-x),\quad\psi_R(t,x)\to\psi_L(t,-x).
\end{align}
The genuine parity transformation $P$ of the fermionic theory with $\Pin^-$ structures satisfies $P^2=(-1)^F$, so it can be defined as $P=\tilde{P}(-1)^{F_L}$:
\begin{equation}
    P:\psi_L(t,x)\to-\psi_R(t,-x),\quad\psi_R(t,x)\to\psi_L(t,-x).
    \label{eq:pin-parity}
\end{equation}

\subsection{Sectors of Majorana CFT}
We consider a circle with $x\sim x+2\pi$. We can impose different boundary conditions on $\psi_L,\psi_R$, so there are four boundary conditions: NSNS, RR, NSR and RNS. We consider only the NSNS and RR boundary conditions.\footnote{Relative to the NSNS boundary condition, the RR, NSR, RNS boundary conditions are obtained by inserting $(-1)^F,(-1)^{F_R},(-1)^{F_L}$ defects, respectively. Since we are interested in $(-1)^F$, we do not consider the NSR and RNS boundary conditions.}

We begin with the NSNS boundary condition. The mode expansions of $\psi_L,\psi_R$ are
\begin{equation}
    \psi_L(0,x)=\sum_{r\in\ZZ+1/2}\psi_{L,r}e^{irx},\quad\psi_R(0,x)=\sum_{r\in\ZZ+1/2}\psi_{R,r}e^{-irx}.
\end{equation}
The ground state $\ket{1}$ is defined as the state annihilated by all modes with $r>0$:
\begin{equation}
    \psi_{L,r}\ket{1}=\psi_{R,r}\ket{1}=0,\quad r>0.
\end{equation}
$\ket{1}$ corresponds to the identity operator $1$. The other primary fields $\psi_L,\psi_R,\psi_L\psi_R$ correspond to the states
\begin{align}
    \psi_{L,-1/2}\ket{1},\quad\psi_{R,-1/2}\ket{1},\quad\psi_{L,-1/2}\psi_{R,-1/2}\ket{1},
\end{align}
respectively. The ground state $\ket{1}$ is assumed to be $P=+1$, and the parity eigenstates are
\begin{align}
    (\psi_{L,-1/2}\pm i\psi_{R,-1/2})\ket{1}&:\ P=\pm i,\\
    \psi_{L,-1/2}\psi_{R,-1/2}\ket{1}&:\ P=+1.
\end{align}

Next, we consider the RR boundary condition. The mode expansions of $\psi_L,\psi_R$ are
\begin{equation}
    \psi_L(0,x)=\sum_{n\in\ZZ}\psi_{L,n}e^{inx},\quad\psi_R(0,x)=\sum_{n\in\ZZ}\psi_{R,n}e^{-inx}.
\end{equation}
Unlike the NS boundary condition, there are fermion zero modes $\psi_{L,0},\psi_{R,0}$. Their anti-commutation relations are
\begin{equation}
    \{\psi_{L,0},\psi_{L,0}\}=\{\psi_{R,0},\psi_{R,0}\}=1,\quad\{\psi_{L,0},\psi_{R,0}\}=0.
\end{equation}
These anti-commutation relations cannot be realized on a one-dimensional representation, so the ground space is two-dimensional. We choose a basis for the ground states so that $\psi_{L,0},\psi_{R,0}$ are represented as $\frac{1}{\sqrt{2}}\sigma^x,\frac{1}{\sqrt{2}}\sigma^z$, respectively.\footnote{Our choice of the basis follows the convention of \cite{Seiberg:2023cdc}.} We can take $(-1)^F$ as $\sigma^y$. The relations
\begin{equation}
    \label{eq:parity_RR}
    P^2=(-1)^F,\quad P\psi_{L,0}=-\psi_{R,0}P,\quad P\psi_{R,0}=\psi_{L,0}P
\end{equation}
are satisfied with
\begin{equation}
    P=\frac{1+i}{2}\begin{pmatrix}
        1 & -1\\
        1 & 1
    \end{pmatrix}.
\end{equation}
The eigenstates of the fermion parity are
\begin{align}
    \ket\mu=\begin{pmatrix}
        1\\ i
    \end{pmatrix}&:\ (-1)^F=+1,\\
    \ket\sigma=\begin{pmatrix}
        1\\-i
    \end{pmatrix}&:\ (-1)^F=-1,
\end{align}
and these states are $P=+1,+i$, respectively. The sectors of the Majorana CFT are summarized in Table~\ref{table:states_majorana}.

\begin{table}[h]
\centering
\begin{tabular}{c|c|c}
    & NSNS & RR \\ \hline
    $P=+1$ & $\ket{1},\psi_{L,-1/2}\psi_{R,-1/2}\ket{1}$ & $\ket\mu$ \\
    $P=-1$ & & \\
    $P=+i$ & $(\psi_{L,-1/2}+i\psi_{R,-1/2})\ket{1}$ & $\ket\sigma$ \\
    $P=-i$ & $(\psi_{L,-1/2}-i\psi_{R,-1/2})\ket{1}$ &
\end{tabular}
\caption{States of the Majorana CFT.}
\label{table:states_majorana}
\end{table}

\subsection{Redefinition of parity}
In the last section, we showed that $S_1^F$ exchanges the $P=+i$ and $P=-i$ sectors. This action can be interpreted as the redefinition of the parity. That is, the operator
\begin{equation}
    P'=P(-1)^F
\end{equation}
satisfies $P'^2=(-1)^F$, and the $P=\pm i$ sectors correspond to the $P'=\mp i$ sectors. $O^F$ acts on $\mathcal{H}_\mathrm{NSNS}$ in the same way as $S_1^F$, so there is no further freedom to redefine the parity in $\mathcal{H}_\mathrm{NSNS}$.\footnote{Redefinition $P\to-P$ preserves the relation $P^2=(-1)^F$, but it changes the eigenvalue of the ground state $\ket{1}$ from $P=+1$ to $P=-1$.}

As in Table~\ref{table:states_RR}, $O^F S_1^F$ acts on the eigenvalues of $P$ by multiplying $-i$ in $\mathcal{H}_\mathrm{RR}$, so this corresponds to the redefinition $P\to P'=-iP$ in $\mathcal{H}_\mathrm{RR}$. Similarly, $(O^F S_1^F)^2,(O^F S_1^F)^3$ corresponds to the redefinition $P\to P'=-P,+iP$, respectively.

Actually, there is no further freedom of redefinition in $\mathcal{H}_\mathrm{RR}$. Our choice $\psi_{L,0}=\frac{1}{\sqrt{2}}\sigma^x,\psi_{R,0}=\frac{1}{\sqrt{2}}\sigma^z$ and the relation
\begin{equation}
    (-1)^F\psi_{L,0}=\psi_{L,0}(-1)^F,\quad (-1)^F\psi_{R,0}=\psi_{R,0}(-1)^F
\end{equation}
fix $(-1)^F=\pm\sigma^y$. The relation $P^2=(-1)^F$ implies that $P$ commutes with $\sigma^y$ and
\begin{equation}
    P=aI+b\sigma^y,
\end{equation}
where $I$ is the identity matrix and $a,b$ are complex numbers. From the relations $P\psi_{L,0}=-\psi_{R,0}P,P\psi_{R,0}=\psi_{L,0}P$, we obtain $a=ib$. $P^2=(-1)^F$ determines $a,b$ and there are just four candidates:
\begin{align}
    P=\pm\frac{1+i}{2}\begin{pmatrix}
        1 & -1\\
        1 & 1
    \end{pmatrix},\quad (-1)^F=\sigma^y,\\
    P=\pm\frac{1-i}{2}\begin{pmatrix}
        1 & -1\\
        1 & 1
    \end{pmatrix},\quad (-1)^F=-\sigma^y,
\end{align}
which correspond to $1,O^FS_1^F,(O^FS_1^F)^2,(O^FS_1^F)^3$. Therefore, we conclude that the actions of $O^F$ and $S_1^F$ are precisely the redefinition of $P$ in the Majorana CFT.

\subsection{Correspondence to Ising CFT}
\paragraph{Correspondence on a lattice} Bosonization of the Majorana CFT is the Ising CFT. We briefly review this correspondence in a lattice formulation. A detailed discussion can be found in \cite{Seiberg:2023cdc, Shao:2023gho}.

A lattice counterpart of the Majorana CFT is the Majorana chain. Consider $L=2N$ real Majorana fermions $\chi_l$ with the anti-commutation relation
\begin{equation}
    \{\chi_l,\chi_{l'}\}=2\delta_{l,l'},
\end{equation}
and two Hamiltonians
\begin{equation}
    H_\pm=i\sum_{l=1}^{L-1}\chi_l\chi_{l+1}\pm i\chi_1\chi_L.
\end{equation}
In the continuum limit, $H_+$ and $H_-$ flow to the Majorana CFT with the RR and NSNS boundary conditions, respectively.

Bosonization of the Majorana chain is the Jordan-Wigner transformation:
\begin{align}
    \chi_{2j-1}=\left(\prod_{j'=1}^{j-1}\sigma_{j'}^x\right)\sigma^z_j,\quad\chi_{2j}=\left(\prod_{j'=1}^{j-1}\sigma_{j'}^x\right)\sigma^y_j,
\end{align}
where $j=1,\dots,N$ and $\sigma^a_j$ are the Pauli matrices. The Hamiltonians can be expressed as
\begin{equation}
    H_\pm=-\sum_{j=1}^N\sigma_j^x-\sum_{j=1}^{N-1}\sigma^z_j\sigma^z_{j+1}\pm(-1)^F\sigma^z_N\sigma^z_1,
\end{equation}
where $(-1)^F=\prod_{j=1}^N\sigma_j^x$. The Hilbert spaces with $H_+,H_-$ are denoted $\mathcal{H}_\text{RR},\mathcal{H}_\text{NSNS}$, respectively. We define an involution $\tilde g$ on $\mathcal{H}_\text{NSNS}\oplus\mathcal{H}_\text{RR}$ as
\begin{equation}
    \tilde g=\begin{pmatrix}
        1 & 0\\ 0 & -1
    \end{pmatrix},
\end{equation}
and decompose $\mathcal{H}_\text{NSNS}\oplus\mathcal{H}_\text{RR}$ by the eigenvalues of $\tilde g(-1)^F$. In the $\tilde g(-1)^F=+1$ eigenspace, the Hamiltonian is
\begin{align}
    H=-\sum_{j=1}^N X_j-\sum_{j=1}^N Z_jZ_{j+1},\quad X_j=\begin{pmatrix}
        \sigma_j^x & 0 \\ 0 & \sigma_j^x
    \end{pmatrix},\quad Z_j=\begin{pmatrix}
        0 & \sigma_j^z\\ \sigma_j^z & 0
    \end{pmatrix}.
\end{align}
This is the Hamiltonian of the transverse-field Ising model, which flows to the Ising CFT in the continuum limit. The Hamiltonian in the $\tilde g(-1)^F=+1$ eigenspace is
\begin{equation}
    H^g=\sum_{j=1}^N X_j-\sum_{j=1}^{N-1} Z_jZ_{j+1}+Z_N Z_1,
\end{equation}
which can be interpreted as the $\ZZ_2$-twisted Hamiltonian. Thus, the Hilbert spaces of the Ising model are
\begin{equation}
    \mathcal{H}_\text{untwisted}=\mathcal{H}_\text{NSNS}^+\oplus\mathcal{H}_\text{R}^-,\quad \mathcal{H}_\text{twisted}=\mathcal{H}_\text{NSNS}^-\oplus\mathcal{H}_\text{R}^+,
\end{equation}
where $\pm$ denotes the eigenvalue of $(-1)^F$.

We have a $\ZZ_2$ symmetry in the transverse-field Ising model:
\begin{align}
    g&=\tilde{g}|_{\mathcal{H}_\text{untwisted}}=(-1)^F|_{\mathcal{H}_\text{untwisted}}\quad \text{in}\ \mathcal{H}_\text{untwisted},\\
    g&=\tilde{g}|_{\mathcal{H}_\text{twisted}}=-(-1)^F|_{\mathcal{H}_\text{twisted}}\quad \text{in}\ \mathcal{H}_\text{twisted},
\end{align}
so we can decompose the Hilbert spaces $\mathcal{H}_\text{untwisted},\mathcal{H}_\text{twisted}$ into eigenspaces of $g$:
\begin{align}
    \mathcal{H}_\text{untwisted}^+&=\mathcal{H}_\text{NSNS}^+,& \mathcal{H}_\text{untwisted}^-&=\mathcal{H}_\text{RR}^-,\nonumber\\
    \mathcal{H}_\text{twisted}^+&=\mathcal{H}_\text{RR}^+,& \mathcal{H}_\text{twisted}^-&=\mathcal{H}_\text{NSNS}^-.
\end{align}
We can define the parity transformation on a lattice, but we do not delve into it.

In the standard convention, the sectors of the Ising CFT are taken as in Table~\ref{table:sectors_ising}. We can see that
\begin{equation}
    \text{Ising CFT}=\Fer^{-1}O^FS_1^F(\text{Majorana CFT}).
    \label{eq:Ising_Majorana}
\end{equation}
The reason $O^F S_1^F$ appears here is that the fermionization $\Fer$ is defined so that the form of $S_1^F$ is simple.

\begin{table}[H]
\centering
\begin{tabular}{c|c|c|c}
    \multicolumn{2}{c|}{} & untwisted & twisted \\ \hline
    \multirow{2}{2em}{even} & $P=+1$ & $\ket{1},\ket{\epsilon}$ & $\ket\mu$ \\
    & $P=-1$ & & \\ \hline
    \multirow{2}{2em}{odd} & $P=+1$ & $\ket{\sigma}$ & $\ket{\psi_L}+\ket{\psi_R}$ \\
    & $P=-1$ & & $\ket{\psi_L}-\ket{\psi_R}$
\end{tabular}
\caption{Sectors of the Ising CFT.}
\label{table:sectors_ising}
\end{table}

\paragraph{Correspondence of characters} Let us consider this correspondence in terms of characters. We can easily calculate the torus partition function of the Majorana CFT:\footnote{See, e.g.~\cite{DiFrancesco:1997nk}.}
\begin{align}
    Z_{T_F}[T^2;\eta_0]&=\Tr_{\mathcal{H}_\mathrm{NSNS}}[e^{-lH}]=\left|q^{-1/48}\prod_{r\in\ZZ_{\ge0}+1/2}(1+q^r)\right|^2=\left|\frac{\theta_3(\tau)}{\eta(\tau)}\right|,\\
    Z_{T_F}[T^2;\eta_0+x]&=\Tr_{\mathcal{H}_\mathrm{NSNS}}[(-1)^Fe^{-lH}]=\left|q^{-1/48}\prod_{r\in\ZZ_{\ge0}+1/2}(1-q^r)\right|^2=\left|\frac{\theta_4(\tau)}{\eta(\tau)}\right|,\\
    Z_{T_F}[T^2;\eta_0+y]&=\Tr_{\mathcal{H}_\mathrm{RR}}[e^{-lH}]=2\left|q^{1/24}\prod_{n=1}^\infty(1+q^n)\right|^2=\left|\frac{\theta_2(\tau)}{\eta(\tau)}\right|,\\
    Z_{T_F}[T^2;\eta_0+x+y]&=\Tr_{\mathcal{H}_\mathrm{RR}}[(-1)^Fe^{-lH}]=0,
\end{align}
where $\tau=\frac{il}{2\pi},q=e^{2\pi i\tau}$ and
\begin{align}
    H&=\sum_{r\in\ZZ_{\ge0}+1/2}r(\psi_{L,-r}\psi_{L,r}+\psi_{R,-r}\psi_{R,r})-\frac{1}{24}&\text{in }\mathcal{H}_\mathrm{NSNS},\\
    H&=\sum_{n=1}^\infty n(\psi_{L,-n}\psi_{L,n}+\psi_{R,-n}\psi_{R,n})+\frac{1}{12}&\text{in }\mathcal{H}_\mathrm{RR}.
\end{align}
The $\theta$ functions and the $\eta$ function are
\begin{equation}
    \begin{gathered}
        \theta_2(\tau)=2q^{1/8}\prod_{n=1}^\infty(1-q^n)(1+q^n)^2,\quad
        \theta_3(\tau)=\prod_{n=1}^\infty(1-q^n)(1+q^{n-1/2})^2,\\
        \theta_4(\tau)=\prod_{n=1}^\infty(1-q^n)(1-q^{n-1/2})^2,\quad\eta(\tau)=q^{1/24}\prod_{n=1}^\infty(1-q^n).
    \end{gathered}
\end{equation}
The Klein bottle partition functions are calculated in a similar way. Since the parity $P$ acts on the fields as in~\eqref{eq:pin-parity}, only states with equal numbers of excitations in the left and right modes contribute to the trace:
\begin{align}
    Z_{T_F}[K;\eta_0']&=\Tr_{\mathcal{H}_\mathrm{NSNS}}[Pe^{-lH}]=q^{-1/24}\prod_{r\in\ZZ_{\ge0}+1/2}(1+q^{2r})=\sqrt{\frac{\theta_3(2\tau)}{\eta(2\tau)}},\\
    Z_{T_F}[K;\eta_0'+x]&=\Tr_{\mathcal{H}_\mathrm{NSNS}}[P^3e^{-lH}]=\sqrt{\frac{\theta_3(2\tau)}{\eta(2\tau)}},\\
    Z_{T_F}[K;\eta_0'+y]&=\Tr_{\mathcal{H}_\mathrm{RR}}[Pe^{-lH}]=q^{1/12}(1+i)\prod_{n=1}^\infty(1+q^{2n})=\frac{1+i}{\sqrt2}\sqrt{\frac{\theta_2(2\tau)}{\eta(2\tau)}},\label{eq:zeromode_trace}\\
    Z_{T_F}[K;\eta_0'+x+y]&=\Tr_{\mathcal{H}_\mathrm{RR}}[P^3e^{-lH}]=\frac{1-i}{\sqrt2}\sqrt{\frac{\theta_2(2\tau)}{\eta(2\tau)}}.
\end{align}
The factor $(1+i)$ appears in~\eqref{eq:zeromode_trace} because $\ket{\mu},\ket{\sigma}$ have eigenvalues $P=+1,+i$, respectively. Bosonization gives the torus partition functions of the Ising CFT:
\begin{align}
    Z_{\mathrm{Fer}^{-1}(T_F)}[T^2;0]&=|\chi_0(\tau)|^2+|\chi_{1/2}(\tau)|^2+|\chi_{1/16}(\tau)|^2,\\
    Z_{\mathrm{Fer}^{-1}(T_F)}[T^2;x]&=|\chi_0(\tau)|^2+|\chi_{1/2}(\tau)|^2-|\chi_{1/16}(\tau)|^2,\\
    Z_{\mathrm{Fer}^{-1}(T_F)}[T^2;y]&=|\chi_{1/16}(\tau)|^2+\chi_0(\tau)\overline{\chi_{1/2}(\tau)}+\chi_{1/2}(\tau)\overline{\chi_0(\tau)},\\
    Z_{\mathrm{Fer}^{-1}(T_F)}[T^2;x+y]&=|\chi_{1/16}(\tau)|^2-\chi_0(\tau)\overline{\chi_{1/2}(\tau)}-\chi_{1/2}(\tau)\overline{\chi_0(\tau)},
\end{align}
where $\chi_0,\chi_{1/2},\chi_{1/16}$ are the Virasoro characters:
\begin{equation}
    \begin{gathered}
    \chi_0(\tau)=\frac{1}{2}\left(\sqrt{\frac{\theta_3(\tau)}{\eta(\tau)}}+\sqrt{\frac{\theta_4(\tau)}{\eta(\tau)}}\right),\quad \chi_{1/2}(\tau)=\frac{1}{2}\left(\sqrt{\frac{\theta_3(\tau)}{\eta(\tau)}}-\sqrt{\frac{\theta_4(\tau)}{\eta(\tau)}}\right),\\
    \chi_{1/16}(\tau)=\sqrt\frac{\theta_2(\tau)}{2\eta(\tau)}.
    \end{gathered}
\end{equation}
The Klein bottle partition functions are
\begin{align}
    Z_{\mathrm{Fer}^{-1}(T_F)}[K;0]&=\chi_0(2\tau)+\chi_{1/2}(2\tau)+\chi_{1/16}(2\tau),\\
    Z_{\mathrm{Fer}^{-1}(T_F)}[K;x]&=\chi_0(2\tau)+\chi_{1/2}(2\tau)-\chi_{1/16}(2\tau),\\
    Z_{\mathrm{Fer}^{-1}(T_F)}[K;y]&=-\chi_{1/16}(2\tau),\\
    Z_{\mathrm{Fer}^{-1}(T_F)}[K;x+y]&=-\chi_{1/16}(2\tau).
\end{align}
The minus sign in $Z_{\mathrm{Fer}^{-1}(F)}[K;y]=\Tr_{\mathcal{H}^g}[Pe^{-lH}]$ means that there is a primary state with $P=-1$ in the twisted Hilbert space. $S_1^B$ flips the signs of $Z_{\mathrm{Fer}^{-1}(F)}[K;y],Z_{\mathrm{Fer}^{-1}(F)}[K;x+y]$ and we get the Klein bottle partition functions of the Ising CFT. This is consistent with the relation~\eqref{eq:Ising_Majorana}.

\subsection*{Acknowledgements}
The authors would like to thank Y.~Tachikawa and T.~Ando for many fruitful discussions and for valuable comments on the draft.

This research was supported in part by the Forefront Physics and Mathematics Program to Drive Transformation (FoPM), a World-leading Innovative Graduate Study (WINGS) Program at the University of Tokyo.

%% file: 6.tex
\appendix
\section{Quadratic refinement}
\label{app}
\subsection{Orientable case}
\paragraph{Spin structure}
Let $X$ be a closed oriented manifold.
A spin structure on $X$ can be described as a trivialization of the second Stiefel--Whitney class\footnote{If $X$ is non-orientable, this definition corresponds to $\Pin^+$ structures.}
\begin{equation}
w_2 \in H^2(X;\mathbb{Z}_2),
\end{equation}
that is, a cochain $\eta \in C^1(X;\mathbb{Z}_2)$ such that
\begin{equation}
\delta \eta = w_2.
\end{equation}

The set of spin structures on $X$ forms a torsor over $H^1(X;\mathbb{Z}_2)$.
Indeed, let $\eta$ and $\eta'$ be two spin structures. Then
\begin{equation}
\delta(\eta - \eta') = \delta\eta - \delta\eta'
= w_2 - w_2 = 0,
\end{equation}
and hence $\eta - \eta'$ is a $1$-cocycle:
\begin{equation}
\eta - \eta' \in Z^1(X;\mathbb{Z}_2).
\end{equation}
Moreover, two such differences that differ by a coboundary,
\begin{equation}
\eta - \eta' \sim \eta - \eta' + \delta\lambda,
\qquad
\lambda \in C^0(X;\mathbb{Z}_2),
\end{equation}
are identified. Therefore, the difference $\eta - \eta'$ defines a cohomology class in $H^1(X;\mathbb{Z}_2).$
Consequently, the set of spin structures on $X$ is naturally a torsor over
$H^1(X;\mathbb{Z}_2)$.

\paragraph{Quadratic refinement}Here we focus on closed two-dimensional manifolds.
Let $\Sigma$ be a closed oriented surface.
Given the canonical pairing
\begin{equation}
H^1(\Sigma;\mathbb{Z}_2)\times H^1(\Sigma;\mathbb{Z}_2)
\to \mathbb{Z}_2,
\qquad
(a,b)\mapsto \int_\Sigma a\cup b ,
\end{equation}
we define a map
\begin{equation}
Q_\eta \colon H^1(\Sigma;\mathbb{Z}_2) \to \mathbb{Z}_2
\end{equation}
by the condition
\begin{equation}
Q_\eta(a+b)
= Q_\eta(a) + Q_\eta(b) + \int_\Sigma a\cup b .
\end{equation}
Such a map is called a quadratic refinement.
Quadratic refinements are known to be in one-to-one correspondence with spin
structures on $\Sigma$, which is why we label $Q_\eta$ by the spin structure
$\eta$~\cite{Johnson}.

One useful property of quadratic refinements is
\begin{equation}
Q_{\eta+a}(b) = Q_\eta(b) + \int_\Sigma a\cup b ,
\qquad
a,b \in H^1(\Sigma;\mathbb{Z}_2).
\end{equation}
Recall that different choices of spin structures form a torsor over
$H^1(\Sigma;\mathbb{Z}_2)$.
Thus, $\eta + a$ corresponds to another spin structure, and
$Q_{\eta+a}$ is the associated quadratic refinement.

Indeed, defining
\begin{equation}
Q_-(x) := Q_{\eta+a}(x) - Q_\eta(x),
\end{equation}
we find that
\begin{equation}
Q_-(x+y) = Q_-(x) + Q_-(y)
\end{equation}
for all $x,y \in H^1(\Sigma;\mathbb{Z}_2)$, which implies that $Q_-$ defines a linear map.
By the non-degeneracy of the intersection pairing, there exists
$a \in H^1(\Sigma;\mathbb{Z}_2)$ such that
\begin{equation}
Q_-(b) = \int_\Sigma a\cup b 
\end{equation}
for all $b\in H^1(\Sigma;\mathbb{Z}_2)$.

\paragraph{Arf invariant}
Using the quadratic refinement $Q_\eta$, one can define a mod-$2$ index by
\begin{equation}
\exp\!\left(\pi i\,\mathrm{Arf}(\eta)\right)
:= \frac{1}{\lvert H^1(\Sigma;\mathbb{Z}_2)\rvert^{1/2}}
\sum_{a \in H^1(\Sigma;\mathbb{Z}_2)} \mathrm{exp}(\pi i \, Q_\eta(a)) .
\end{equation}
This quantity is a well-known invariant taking values in $\mathbb{Z}_2$,
called the Arf invariant.
Physically, it admits the following interpretations:
\begin{itemize}
  \item the partition function of the Kitaev chain regarded as a topological phase~\cite{Kitaev:2000nmw},
  \item the mod-$2$ index of the Dirac operator.
\end{itemize}

In short, this invariant defines a homomorphism
\begin{equation}
\Omega_2^{\mathrm{Spin}}(\mathrm{pt}) \to \mathbb{Z}_2 ,
\end{equation}
which classifies spin structures into two distinct classes.

Let us derive a basic relation associated with the Arf invariant:
\begin{equation}
\mathrm{Arf}(\eta+a)-\mathrm{Arf}(\eta)=Q_\eta(a),
\qquad a\in H^1(\Sigma;\mathbb{Z}_2),
\end{equation}
where both sides are understood in $\mathbb{Z}_2$.
We compute
\begin{equation}
\begin{aligned}
\exp\!\left(\pi i\bigl(\mathrm{Arf}(\eta+a)-\mathrm{Arf}(\eta)\bigr)\right)
&=\exp(\pi i\,\mathrm{Arf}(\eta+a))\,\exp(-\pi i\,\mathrm{Arf}(\eta))\\
&=\frac{1}{\lvert H^1(\Sigma;\mathbb{Z}_2)\rvert}
\sum_{x,y\in H^1(\Sigma;\mathbb{Z}_2)}
\exp\!\left(\pi i\bigl(Q_{\eta+a}(x)-Q_\eta(y)\bigr)\right).
\end{aligned}
\end{equation}
Set $z:=x+y$, then
\begin{equation}
\begin{aligned}
Q_{\eta+a}(x)-Q_\eta(y)
&=Q_{\eta+a}(x)-Q_\eta(z-x)\\
&=Q_{\eta+a}(x)-\Bigl(Q_\eta(z)+Q_\eta(x)+\int_\Sigma z\cup x\Bigr)\\
&=\bigl(Q_{\eta+a}(x)-Q_\eta(x)\bigr)-Q_\eta(z)-\int_\Sigma z\cup x\\
&=\int_\Sigma a\cup x - Q_\eta(z) - \int_\Sigma z\cup x\\
&= -Q_\eta(z) + \int_\Sigma (a+z)\cup x.
\end{aligned}
\end{equation}
Therefore,
\begin{equation}
\begin{aligned}
\exp\!\left(\pi i\bigl(\mathrm{Arf}(\eta+a)-\mathrm{Arf}(\eta)\bigr)\right)
&=\frac{1}{\lvert H^1(\Sigma;\mathbb{Z}_2)\rvert}
\sum_{x,z\in H^1(\Sigma;\mathbb{Z}_2)}
\exp\!\left(\pi i\Bigl(-Q_\eta(z)+\int_\Sigma (a+z)\cup x\Bigr)\right)\\
&=\frac{1}{\lvert H^1(\Sigma;\mathbb{Z}_2)\rvert}
\sum_{z\in H^1(\Sigma;\mathbb{Z}_2)} (-1)^{q_\eta(z)}
\sum_{x\in H^1(\Sigma;\mathbb{Z}_2)} (-1)^{\int_\Sigma (a+z)\cup x}.
\end{aligned}
\end{equation}
Here we have the following:
\begin{equation}
\sum_{x\in H^1(\Sigma;\mathbb{Z}_2)} (-1)^{\int_\Sigma (a+z)\cup x}
=
\lvert H^1(\Sigma;\mathbb{Z}_2)\rvert\,\delta_{z,a}.
\end{equation}
Hence only $z=a$ contributes, and we obtain
\begin{equation}
\exp\!\left(\pi i\bigl(\mathrm{Arf}(\eta+a)-\mathrm{Arf}(\eta)\bigr)\right)
=(-1)^{Q_\eta(a)}=\exp(\pi i\,Q_\eta(a)).
\end{equation}
This proves
\begin{equation}
\mathrm{Arf}(\eta+a)-\mathrm{Arf}(\eta)=Q_\eta(a)\in\mathbb{Z}_2,
\end{equation}
as claimed.

\subsection{Non-orientable case}
\label{Nonorientable abk}
\paragraph{$\Pin^-$ structure}
Let $X$ be a closed (possibly non-orientable) manifold.
A $\Pin^-$ structure on $X$ is a trivialization of
\begin{equation}
w_2 + w_1^2 \in H^2(X;\mathbb{Z}_2),
\end{equation}
that is, a cochain $\eta \in C^1(X;\mathbb{Z}_2)$ such that
\begin{equation}
\delta \eta = w_2 + w_1^2 .
\end{equation}
As in the discussion above, the set of $\Pin^-$ structures on $X$
forms a torsor over $H^1(X,\mathbb{Z}_2)$.

Note that on any two-dimensional manifold one has
\begin{equation}
w_2 + w_1^2 = 0 .
\end{equation}
Therefore, every two-dimensional manifold admits $\Pin^-$ structures.
\paragraph{quadratic refinement}
Here we focus on closed two-dimensional manifolds.
Let $\Sigma$ be a closed (possibly non-orientable) surface.
Given the canonical pairing
\begin{equation}
H^1(\Sigma;\mathbb{Z}_2)\times H^1(\Sigma;\mathbb{Z}_2)
\to \mathbb{Z}_2,
\qquad
(a,b)\mapsto \int_\Sigma a\cup b ,
\end{equation}
a $\mathrm{Pin}^-$ structure $\eta$ on $\Sigma$ determines a map
\begin{equation}
q_\eta \colon H^1(\Sigma;\mathbb{Z}_2) \to \mathbb{Z}_4
\end{equation}
characterized by
\begin{equation}
q_\eta(a+b)
= q_\eta(a) + q_\eta(b) + 2\int_\Sigma a\cup b.
\end{equation}
Such a map is called a $\mathbb{Z}_4$-valued quadratic refinement of the
intersection pairing.
On a closed surface, $\mathbb{Z}_4$-valued quadratic refinements are in
one-to-one correspondence with $\mathrm{Pin}^-$ structures on $\Sigma$~\cite{Kirby_Taylor_1991}.

Note that the $\mathbb{Z}_2$-valued quadratic refinement defined for spin structures is a special case of this $\mathbb{Z}_4$-valued quadratic refinement. We denote $\ZZ_2$-valued (resp.~$\ZZ_4$-valued) quadratic refinements by $Q$ (resp.~$q$). When $\Sigma$ is oriented, they are related by $q=2Q$.

One can also check the following identity:
\begin{equation}
q_{\eta+a}(b)
= q_\eta(b) + 2\int_\Sigma a\cup b ,
\qquad
a,b \in H^1(\Sigma;\mathbb{Z}_2),
\end{equation}
where the equality is understood in $\mathbb{Z}_4$.

\paragraph{ABK invariant}
Using the $\mathbb{Z}_4$-valued quadratic refinement $q_\eta$ associated with a
$\mathrm{Pin}^-$ structure, one can define a bordism invariant by
\begin{equation}
\exp\!\left(\frac{\pi i}{4}\,\mathrm{ABK}(\eta)\right)
:= \frac{1}{\lvert H^1(\Sigma;\mathbb{Z}_2)\rvert^{1/2}}
\sum_{a \in H^1(\Sigma;\mathbb{Z}_2)} \exp\!\left(\frac{\pi i}{2}\, q_\eta(a)\right) .
\end{equation}
This quantity is a well-known invariant taking values in $\mathbb{Z}_8$, called the
Arf--Brown--Kervaire (ABK) invariant.

Physically, it admits the following interpretations:
\begin{itemize}
  \item the partition function of the time-reversal-invariant Kitaev Majorana chain~\cite{Fidkowski:2010jmn},
  \item the mod-$8$ index of the Dirac operator.
\end{itemize}

In short, this invariant defines a homomorphism
\begin{equation}
\Omega_2^{\mathrm{Pin}^-}(\mathrm{pt}) \to \mathbb{Z}_8 ,
\end{equation}
which classifies $\mathrm{Pin}^-$ structures into eight distinct bordism
classes.

One can also check the following identity:
\begin{equation}
\frac{1}{2}\left(\ABK(\eta+a)-\ABK(\eta)\right)=q_\eta(a)+2\int a\cup a= -q_\eta(a)\quad\in\mathbb{Z}_4.\label{abk and quad}
\end{equation}
To show this, we compute
\begin{equation}
  \exp\!\left(\frac{\pi i}{4}\,
    \bigl(\mathrm{ABK}(\eta+a)-\mathrm{ABK}(\eta)\bigr)\right)
  =
  \frac{1}{\lvert H^1(\Sigma;\mathbb{Z}_2)\rvert}
  \sum_{b,c \in H^1(\Sigma;\mathbb{Z}_2)}
  \exp\!\left(\frac{\pi i}{2}\,
    \bigl(q_{\eta+a}(b)-q_\eta(c)\bigr)\right).
\end{equation}
Now define \(d = b - c\). Then we have
\begin{equation}
\begin{aligned}
  q_{\eta+a}(b) - q_\eta(c)
  &= q_\eta(b) + 2\!\int a \cup b - q_\eta(c) \\
  &= q_\eta(d+c) + 2\!\int a \cup (d+c) - q_\eta(c) \\
  &= q_\eta(d)
     + 2\!\int a \cup d
     + 2\!\int (a+d) \cup c .
\end{aligned}
\end{equation}
Performing the sum over \(c\), we obtain
\begin{equation}
  \exp\!\left(\frac{\pi i}{4}\,
    \bigl(\mathrm{ABK}(\eta+a)-\mathrm{ABK}(\eta)\bigr)\right)
  =
  \exp\!\left(\frac{\pi i}{2}\, q_\eta(a)\right)
  (-1)^{\int a \cup a}.
\end{equation}
By the definition of a quadratic refinement, we have
\begin{equation}
  2 q_\eta(a) = 2 \int a \cup a .
\end{equation}
Therefore,
\begin{equation}
  \exp\!\left(\frac{\pi i}{4}\,
    \bigl(\mathrm{ABK}(\eta+a)-\mathrm{ABK}(\eta)\bigr)\right)
  =
  \exp\!\left(\frac{3\pi i}{2}\, q_\eta(a)\right)
  =
  \exp\!\left(-\frac{\pi i}{2}\, q_\eta(a)\right),
\end{equation}
which is the desired relation.

%% file: ref.bib
@article{Hsieh:2020uwb,
    author = "Hsieh, Chang-Tse and Nakayama, Yu and Tachikawa, Yuji",
    title = "{Fermionic minimal models}",
    eprint = "2002.12283",
    archivePrefix = "arXiv",
    primaryClass = "cond-mat.str-el",
    reportNumber = "IPMU-20-0008, RUP-20-5",
    doi = "10.1103/PhysRevLett.126.195701",
    journal = "Phys. Rev. Lett.",
    volume = "126",
    number = "19",
    pages = "195701",
    year = "2021"
}

@article{Spieler:2025hyr,
    author = "Spieler, Ryan C.",
    title = "{Exploring two dimensional ${\mathbb{Z}}_{2}$ invariant phases with time reversal symmetry and their transitions with topological operations}",
    eprint = "2504.20021",
    archivePrefix = "arXiv",
    primaryClass = "cond-mat.str-el",
    doi = "10.1007/JHEP10(2025)069",
    journal = "JHEP",
    volume = "10",
    pages = "069",
    year = "2025"
}

@article{Karch:2019lnn,
    author = "Karch, Andreas and Tong, David and Turner, Carl",
    title = "{A Web of 2d Dualities: ${\bf Z}_2$ Gauge Fields and Arf Invariants}",
    eprint = "1902.05550",
    archivePrefix = "arXiv",
    primaryClass = "hep-th",
    doi = "10.21468/SciPostPhys.7.1.007",
    journal = "SciPost Phys.",
    volume = "7",
    pages = "007",
    year = "2019"
}

@article{BoyleSmith:2024qgx,
    author = "Boyle Smith, Philip and Zheng, Yunqin",
    title = "{Backfiring Bosonisation}",
    eprint = "2403.03953",
    archivePrefix = "arXiv",
    primaryClass = "hep-th",
    month = "3",
    year = "2024"
}

@article{Kobayashi:2022qhc,
    author = "Kobayashi, Ryohei",
    title = "{Fermionic topological phases and bosonization in higher dimensions}",
    doi = "10.1093/ptep/ptab110",
    journal = "PTEP",
    volume = "2022",
    number = "4",
    pages = "04A105",
    year = "2022"
}

@article{Kitaev:2000nmw,
    author = "Kitaev, Alexei",
    title = "{Unpaired Majorana fermions in quantum wires}",
    eprint = "cond-mat/0010440",
    archivePrefix = "arXiv",
    doi = "10.1070/1063-7869/44/10S/S29",
    journal = "Phys. Usp.",
    volume = "44",
    number = "10S",
    pages = "131--136",
    year = "2001"
}

@article{Kirby_Taylor_1991, 
place={Cambridge}, 
journal={London Mathematical Society Lecture Note Series}, 
title={Pin structures on low-dimensional manifolds}, 
booktitle={Geometry of Low-Dimensional Manifolds: Symplectic Manifolds and Jones-Witten Theory}, 
publisher={Cambridge University Press}, 
author={Kirby, R.C. and Taylor, L.R.}, 
editor={Donaldson, S. K. and Thomas, C. B.Editors}, 
year={1991}, 
volume={151},
pages={177--242}, 
collection={London Mathematical Society Lecture Note Series},
}

@article{Johnson,
author = {Johnson, Dennis},
title = {Spin Structures and Quadratic forms on Surfaces},
journal = {Journal of the London Mathematical Society},
series=2,
volume = {22},
number = {2},
pages = {365-373},
doi = {10.1112/jlms/s2-22.2.365},
year = {1980}
}

@article{Fidkowski:2010jmn,
    author = "Fidkowski, Lukasz and Kitaev, Alexei",
    title = "{Topological phases of fermions in one dimension}",
    eprint = "1008.4138",
    archivePrefix = "arXiv",
    primaryClass = "cond-mat.str-el",
    doi = "10.1103/PhysRevB.83.075103",
    journal = "Phys. Rev. B",
    volume = "83",
    number = "7",
    pages = "075103",
    year = "2011"
}

@article{Kapustin:2014gma,
    author = "Kapustin, Anton",
    title = "{Bosonic Topological Insulators and Paramagnets: a view from cobordisms}",
    eprint = "1404.6659",
    archivePrefix = "arXiv",
    primaryClass = "cond-mat.str-el",
    month = "4",
    year = "2014"
}

@article{Seiberg:2023cdc,
    author = "Seiberg, Nathan and Shao, Shu-Heng",
    title = "{Majorana chain and Ising model - (non-invertible) translations, anomalies, and emanant symmetries}",
    eprint = "2307.02534",
    archivePrefix = "arXiv",
    primaryClass = "cond-mat.str-el",
    reportNumber = "YITP-SB-2023-14",
    doi = "10.21468/SciPostPhys.16.3.064",
    journal = "SciPost Phys.",
    volume = "16",
    number = "3",
    pages = "064",
    year = "2024"
}

@inproceedings{Shao:2023gho,
    author = "Shao, Shu-Heng",
    title = "{What's Done Cannot Be Undone: TASI Lectures on Non-Invertible Symmetries}",
    booktitle = "{TASI 2023}: {Aspects of Symmetry}",
    eprint = "2308.00747",
    archivePrefix = "arXiv",
    primaryClass = "hep-th",
    reportNumber = "YITP-SB-2023-19",
    year = "2023"
}

@article{Barkeshli:2016mew,
    author = "Barkeshli, Maissam and Bonderson, Parsa and Jian, Chao-Ming and Cheng, Meng and Walker, Kevin",
    title = "{Reflection and time reversal symmetry enriched topological phases of matter: path integrals, non-orientable manifolds, and anomalies}",
    eprint = "1612.07792",
    archivePrefix = "arXiv",
    primaryClass = "cond-mat.str-el",
    doi = "10.1007/s00220-019-03475-8",
    journal = "Commun. Math. Phys.",
    volume = "374",
    number = "2",
    pages = "1021--1124",
    year = "2019"
}

@article{Orii:2025ktu,
    author = "Orii, Ippo",
    title = "{On dimensions of (2+1)D abelian bosonic topological systems on unoriented manifolds}",
    eprint = "2502.13532",
    archivePrefix = "arXiv",
    primaryClass = "hep-th",
    doi = "10.1093/ptep/ptaf056",
    journal = "PTEP",
    volume = "2025",
    pages = "053",
    month = "2",
    year = "2025"
}

@article{Thorngren:2014pza,
    author = "Thorngren, Ryan",
    title = "{Framed Wilson Operators, Fermionic Strings, and Gravitational Anomaly in 4d}",
    eprint = "1404.4385",
    archivePrefix = "arXiv",
    primaryClass = "hep-th",
    doi = "10.1007/JHEP02(2015)152",
    journal = "JHEP",
    volume = "02",
    pages = "152",
    year = "2015"
}

@article{Kapustin:2017jrc,
    author = "Kapustin, Anton and Thorngren, Ryan",
    title = "{Fermionic SPT phases in higher dimensions and bosonization}",
    eprint = "1701.08264",
    archivePrefix = "arXiv",
    primaryClass = "cond-mat.str-el",
    doi = "10.1007/JHEP10(2017)080",
    journal = "JHEP",
    volume = "10",
    pages = "080",
    year = "2017"
}

@article{Turzillo:2023yyr,
    author = "Turzillo, Alex and You, Minyoung",
    title = "{Duality and stacking of bosonic and fermionic SPT phases}",
    eprint = "2311.18782",
    archivePrefix = "arXiv",
    primaryClass = "cond-mat.str-el",
    doi = "10.1007/JHEP10(2024)034",
    journal = "JHEP",
    volume = "10",
    pages = "034",
    year = "2024"
}

@article{Duan:2023ykn,
    author = "Duan, Zhihao and Jia, Qiang and Lee, Sungjay",
    title = "{{\ensuremath{\mathbb{Z}}}$_{N}$ duality and parafermions revisited}",
    eprint = "2309.01913",
    archivePrefix = "arXiv",
    primaryClass = "hep-th",
    reportNumber = "KIAS-P23039",
    doi = "10.1007/JHEP11(2023)206",
    journal = "JHEP",
    volume = "11",
    pages = "206",
    year = "2023"
}

@article{Kulp:2020iet,
    author = "Kulp, Justin",
    title = "{Two More Fermionic Minimal Models}",
    eprint = "2003.04278",
    archivePrefix = "arXiv",
    primaryClass = "hep-th",
    doi = "10.1007/JHEP03(2021)124",
    journal = "JHEP",
    volume = "03",
    pages = "124",
    year = "2021"
}

@article{Gaiotto:2020iye,
    author = "Gaiotto, Davide and Kulp, Justin",
    title = "{Orbifold groupoids}",
    eprint = "2008.05960",
    archivePrefix = "arXiv",
    primaryClass = "hep-th",
    doi = "10.1007/JHEP02(2021)132",
    journal = "JHEP",
    volume = "02",
    pages = "132",
    year = "2021"
}

@article{Apruzzi:2021nmk,
    author = "Apruzzi, Fabio and Bonetti, Federico and Garc{\'\i}a Etxebarria, I{\~n}aki and Hosseini, Saghar S. and Schafer-Nameki, Sakura",
    title = "{Symmetry TFTs from String Theory}",
    eprint = "2112.02092",
    archivePrefix = "arXiv",
    primaryClass = "hep-th",
    doi = "10.1007/s00220-023-04737-2",
    journal = "Commun. Math. Phys.",
    volume = "402",
    number = "1",
    pages = "895--949",
    year = "2023"
}

@article{Ji:2019ugf,
    author = "Ji, Wenjie and Shao, Shu-Heng and Wen, Xiao-Gang",
    title = "{Topological Transition on the Conformal Manifold}",
    eprint = "1909.01425",
    archivePrefix = "arXiv",
    primaryClass = "cond-mat.str-el",
    doi = "10.1103/PhysRevResearch.2.033317",
    journal = "Phys. Rev. Res.",
    volume = "2",
    number = "3",
    pages = "033317",
    year = "2020"
}

@article{Schafer-Nameki:2023jdn,
    author = "Schafer-Nameki, Sakura",
    title = "{ICTP lectures on (non-)invertible generalized symmetries}",
    eprint = "2305.18296",
    archivePrefix = "arXiv",
    primaryClass = "hep-th",
    doi = "10.1016/j.physrep.2024.01.007",
    journal = "Phys. Rept.",
    volume = "1063",
    pages = "1--55",
    year = "2024"
}

@article{Freed:2022qnc,
    author = "Freed, Daniel S. and Moore, Gregory W. and Teleman, Constantin",
    title = "{Topological symmetry in quantum field theory}",
    eprint = "2209.07471",
    archivePrefix = "arXiv",
    primaryClass = "hep-th",
    month = "9",
    year = "2022"
}

@article{Kaidi:2023maf,
    author = "Kaidi, Justin and Nardoni, Emily and Zafrir, Gabi and Zheng, Yunqin",
    title = "{Symmetry TFTs and anomalies of non-invertible symmetries}",
    eprint = "2301.07112",
    archivePrefix = "arXiv",
    primaryClass = "hep-th",
    doi = "10.1007/JHEP10(2023)053",
    journal = "JHEP",
    volume = "10",
    pages = "053",
    year = "2023"
}

@article{Kobayashi:2019xxg,
    author = "Kobayashi, Ryohei",
    title = "{Pin TQFT and Grassmann integral}",
    eprint = "1905.05902",
    archivePrefix = "arXiv",
    primaryClass = "cond-mat.str-el",
    doi = "10.1007/JHEP12(2019)014",
    journal = "JHEP",
    volume = "12",
    pages = "014",
    year = "2019"
}

@article{Gu:2012ib,
    author = "Gu, Zheng-Cheng and Wen, Xiao-Gang",
    title = "{Symmetry-protected topological orders for interacting fermions: Fermionic topological nonlinear {\ensuremath{\sigma}} models and a special group supercohomology theory}",
    eprint = "1201.2648",
    archivePrefix = "arXiv",
    primaryClass = "cond-mat.str-el",
    doi = "10.1103/PhysRevB.90.115141",
    journal = "Phys. Rev. B",
    volume = "90",
    number = "11",
    pages = "115141",
    year = "2014"
}

@article{Bhardwaj:2016clt,
    author = "Bhardwaj, Lakshya and Gaiotto, Davide and Kapustin, Anton",
    title = "{State sum constructions of spin-TFTs and string net constructions of fermionic phases of matter}",
    eprint = "1605.01640",
    archivePrefix = "arXiv",
    primaryClass = "cond-mat.str-el",
    doi = "10.1007/JHEP04(2017)096",
    journal = "JHEP",
    volume = "04",
    pages = "096",
    year = "2017"
}

@article{Wen:2024udn,
    author = "Wen, Rui and Ye, Weicheng and Potter, Andrew C.",
    title = "{Topological holography for fermions}",
    eprint = "2404.19004",
    archivePrefix = "arXiv",
    primaryClass = "cond-mat.str-el",
    month = "4",
    year = "2024"
}

@article{Bhardwaj:2024ydc,
    author = "Bhardwaj, Lakshya and Inamura, Kansei and Tiwari, Apoorv",
    title = "{Fermionic non-invertible symmetries in (1+1)d: Gapped and gapless phases, transitions, and symmetry TFTs}",
    eprint = "2405.09754",
    archivePrefix = "arXiv",
    primaryClass = "hep-th",
    doi = "10.21468/SciPostPhys.18.6.194",
    journal = "SciPost Phys.",
    volume = "18",
    number = "6",
    pages = "194",
    year = "2025"
}

@article{Gaiotto:2014kfa,
    author = "Gaiotto, Davide and Kapustin, Anton and Seiberg, Nathan and Willett, Brian",
    title = "{Generalized Global Symmetries}",
    eprint = "1412.5148",
    archivePrefix = "arXiv",
    primaryClass = "hep-th",
    doi = "10.1007/JHEP02(2015)172",
    journal = "JHEP",
    volume = "02",
    pages = "172",
    year = "2015"
}

@book{DiFrancesco:1997nk,
    author = "Di Francesco, P. and Mathieu, P. and Senechal, D.",
    title = "{Conformal Field Theory}",
    doi = "10.1007/978-1-4612-2256-9",
    isbn = "978-0-387-94785-3, 978-1-4612-7475-9",
    publisher = "Springer-Verlag",
    address = "New York",
    series = "Graduate Texts in Contemporary Physics",
    year = "1997"
}
